%% file: 3HDMDelta27.tex
\documentclass[11pt,a4paper]{article}
\usepackage{jheppub}

\usepackage{graphicx}
\usepackage{amssymb}
\usepackage{amsmath}
\usepackage{subcaption}
\usepackage{xspace}
\usepackage{xcolor}
\usepackage[utf8]{inputenc}
\usepackage{dsfont} %%% for identity matrix
\usepackage{subcaption}
\usepackage{soul}

\usepackage{listings}
\usepackage[
]{hyperref}
\usepackage{orcidlink} % must come AFTER hyperref

\input{tex/lst_styles.tex}

\newcommand{\sa}{\texttt{SARAH}\@\xspace}
\newcommand{\hb}{\texttt{HiggsBounds}\@\xspace}
\newcommand{\hs}{\texttt{HiggsSignals}\@\xspace}

\title{
3HDM with $\Delta(27)$ symmetry and its phenomenological consequences
}
\author[a,b]{J. Kalinowski\orcidlink{0000-0001-5618-0141},}
\affiliation[a]{Faculty of Physics, University of Warsaw, Pasteura 5, 02-093 Warsaw, Poland}
\affiliation[b]{CERN, Theoretical Physics Department, Geneva, Switzerland}
\author[c,d]{W. Kotlarski\orcidlink{0000-0002-1191-6343},}
\affiliation[c]{Institut f\"ur Kern- und Teilchenphysik, TU Dresden, Zellescher Weg 19, 01069 Dresden, Germany}
\affiliation[d]{National Centre for Nuclear Research, Pasteura 7, 02-093 Warsaw, Poland}
\author[e]{M. N. Rebelo\orcidlink{0000-0002-8744-5146},}
\author[e]{and I. de Medeiros Varzielas\orcidlink{0000-0003-4996-3484}}
\affiliation[e]{Centro de F\'isica Te\'orica de Part\'iculas, Instituto Superior T\'ecnico,  Universidade de Lisboa, Av.\ Rovisco Pais 1, P-1049-001 Lisboa, Portugal}

\emailAdd{jan.kalinowski@fuw.edu.pl}
\emailAdd{wojciech.kotlarski@ncbj.gov.pl}
\emailAdd{rebelo@tecnico.ulisboa.pt}
\emailAdd{ivo.de@udo.edu}

\input{tex/abstract}

\begin{document}
{\hfill  CERN-TH-2021-182}

\maketitle

\input{tex/introduction}
\input{tex/model}

\input{tex/phenomenology}
\input{tex/conclusions}
\input{tex/acknowledgments}

\begin{appendix}
\input{tex/D27group}

\input{tex/spheno_setup}
\input{tex/decays}
\end{appendix}

\bibliographystyle{JHEP}
\bibliography{bibliography}

\end{document}

%% file: tex/lst_styles.tex
\definecolor{maroon}{cmyk}{0, 0.87, 0.68, 0.32}
\definecolor{halfgray}{gray}{0.55}
\definecolor{slha_frame}{RGB}{207, 207, 207}
\definecolor{slha_bg}{RGB}{247, 247, 247}
\definecolor{slha_red}{RGB}{186, 33, 33}
\definecolor{slha_green}{RGB}{0, 128, 0}
\definecolor{slha_cyan}{RGB}{64, 128, 128}
\definecolor{slha_purple}{RGB}{170, 34, 255}

\definecolor{mathematica_frame}{RGB}{207, 207, 207}
\definecolor{mathematica_bg}{RGB}{247, 247, 247}
\definecolor{mathematica_red}{RGB}{186, 33, 33}
\definecolor{mathematica_green}{RGB}{0, 128, 0}
\definecolor{mathematica_cyan}{RGB}{64, 128, 128}
\definecolor{mathematica_purple}{RGB}{170, 34, 255}

\lstnewenvironment{MIN}[1][]{%
  \renewcommand{\thelstnumber}{In[\arabic{lstnumber}]}
  \lstset{language=MathIn,numbers=left,basicstyle=\ttfamily,#1}%
}{%
}

\lstnewenvironment{MOUT}[1][]{%
  \renewcommand{\thelstnumber}{Out[\arabic{lstnumber}]}
  \lstset{language=MathOut,numbers=left,basicstyle=\ttfamily,#1}%
}{%
}

\usepackage{listings}
\lstdefinelanguage{SLHA}{
    morekeywords={block,Block,BLOCK,decay,Decay,DECAY},%
    sensitive=true,%
    morecomment=[l]\#,%
    morestring=[b]',%
    morestring=[b]",%
    morestring=[s]{'''}{'''},% used for documentation text (mulitiline strings)
    morestring=[s]{"""}{"""},% added by Philipp Matthias Hahn
    morestring=[s]{r'}{'},% `raw' strings
    morestring=[s]{r"}{"},%
    morestring=[s]{r'''}{'''},%
    morestring=[s]{r"""}{"""},%
    morestring=[s]{u'}{'},% unicode strings
    morestring=[s]{u"}{"},%
    morestring=[s]{u'''}{'''},%
    morestring=[s]{u"""}{"""},%
    identifierstyle=\color{black}\ttfamily,
    commentstyle=\color{slha_cyan}\ttfamily,
    stringstyle=\color{slha_red}\ttfamily,
    keepspaces=true,
    showspaces=false,
    showstringspaces=false,
    rulecolor=\color{slha_frame},
    frame=single,
    frameround={t}{t}{t}{t},
    framexleftmargin=6mm,
    numbers=left,
    numberstyle=\tiny\color{halfgray},
    backgroundcolor=\color{slha_bg},
    basicstyle=\footnotesize,
    keywordstyle=\color{slha_green}\ttfamily,
    aboveskip=1.2em,
    belowskip=1.2em,
}

\lstdefinelanguage{MathIn}{
    morekeywords={Simplify,Eigenvalues},%
    emph={Start,InitUnitarity,GetScatteringDiagrams,BuildScatteringMatrix,MakeSPheno},%
    emphstyle={\color{mathematica_purple}},
    sensitive=true,%
    morecomment=[l]\%,%
    morestring=[b]',%
    morestring=[b]",%
    morestring=[s]{'''}{'''},% used for documentation text (mulitiline strings)
    morestring=[s]{"""}{"""},% added by Philipp Matthias Hahn
    morestring=[s]{r'}{'},% `raw' strings
    morestring=[s]{r"}{"},%
    morestring=[s]{r'''}{'''},%
    morestring=[s]{r"""}{"""},%
    morestring=[s]{u'}{'},% unicode strings
    morestring=[s]{u"}{"},%
    morestring=[s]{u'''}{'''},%
    morestring=[s]{u"""}{"""},%
    identifierstyle=\color{black}\ttfamily,
    commentstyle=\color{mathematica_cyan}\ttfamily,
    stringstyle=\color{mathematica_red}\ttfamily,
    keepspaces=true,
    showspaces=false,
    showstringspaces=false,
    rulecolor=\color{mathematica_frame},
    frame=single,
    frameround={t}{t}{t}{t},
    framexleftmargin=10mm,
    numbers=left,
    numberstyle=\tiny\color{halfgray},
    backgroundcolor=\color{mathematica_bg},
    basicstyle=\footnotesize,
    keywordstyle=\color{mathematica_green}\ttfamily,
    aboveskip=1.2em,
    belowskip=1.2em,
}

\lstdefinelanguage{MathOut}{
    morekeywords={Simplify,Eigenvalues},%
    sensitive=true,%
    morecomment=[l]\%,%
    morestring=[b]',%
    morestring=[b]",%
    morestring=[s]{'''}{'''},% used for documentation text (mulitiline strings)
    morestring=[s]{"""}{"""},% added by Philipp Matthias Hahn
    morestring=[s]{r'}{'},% `raw' strings
    morestring=[s]{r"}{"},%
    morestring=[s]{r'''}{'''},%
    morestring=[s]{r"""}{"""},%
    morestring=[s]{u'}{'},% unicode strings
    morestring=[s]{u"}{"},%
    morestring=[s]{u'''}{'''},%
    morestring=[s]{u"""}{"""},%
    identifierstyle=\color{black}\ttfamily,
    commentstyle=\color{mathematica_cyan}\ttfamily,
    stringstyle=\color{mathematica_red}\ttfamily,
    keepspaces=true,
    showspaces=false,
    showstringspaces=false,
    rulecolor=\color{mathematica_frame},
    frame=single,
    frameround={t}{t}{t}{t},
    framexleftmargin=10mm,
    numbers=left,
    numberstyle=\tiny\color{halfgray},
    backgroundcolor=\color{mathematica_bg},
    basicstyle=\footnotesize,
    keywordstyle=\color{mathematica_green}\ttfamily,
    aboveskip=1.2em,
    belowskip=1.2em,
}

\let\origthelstnumber\thelstnumber
\makeatletter
\newcommand*\Suppressnumber{%
  \lst@AddToHook{OnNewLine}{%
    \let\thelstnumber\relax%
     \advance\c@lstnumber-\@ne\relax%
    }%
}

\newcommand*\Reactivatenumber{%
  \lst@AddToHook{OnNewLine}{%
   \let\thelstnumber\origthelstnumber%
   \advance\c@lstnumber\@ne\relax}%
}

%% file: tex/abstract.tex
\abstract{
We perform a comprehensive analysis of a version of the 3-Higgs doublet model whose scalar potential is invariant under a global $\Delta(27)$ discrete symmetry and where the three scalar doublets are chosen to transform as a triplet under this discrete group.
For each of the known tree-level minima we study the mass spectra and use the oblique parameters $STU$ as well as perturbative unitarity to constrain the parameter space of the model.
We then discuss phenomenological consequences of some leading order  flavour mixing quark Yukawa couplings by considering the flavour violation process  $b \to s \gamma$.

We show that perturbative unitarity significantly constrains parameters of the model while, conversely, the beyond the Standard Model contributions to the $b \to s \gamma$ decay are automatically tamed by the symmetry.
}

%% file: tex/introduction.tex
\section{Introduction}

The discovery of the Higgs boson and  the (so far) non-observation of new particles at the LHC have confirmed the Standard
Model as our
paradigm. Only the observation of neutrino mixing requires its extension  in the
leptonic sector.  However, a number of theoretical arguments as well as experimental observations still motivate searches for a more
fundamental theory.
Extensions of the scalar sector of the Standard Model (SM) are particularly well motivated. One of the simplest ones is the well-known two-Higgs doublet model
(2HDM) which, in its different incarnations, has been already extensively studied (see for example 
\cite{Gunion:1989we,Branco:2011iw}).
Models with more Higgs doublets transforming non-trivially
under non-Abelian discrete symmetries have also gained interest, in particular in the context of predicting the observed mixing patterns of leptons or quarks.
That is because, in its most general form, already a three-Higgs doublet model (3HDM) has a large number of free parameters  in the scalar sector alone. Therefore it is particularly appealing from the point of view of model
building to impose additional symmetries in the scalar sector.
Symmetries play an important role in reducing the number of free parameters and,
as a result, they increase the predictive power of the model. Continuous symmetries,
when spontaneously broken, lead to undesired massless Goldstone bosons.
There is however a small list of finite symmetries a 3HDM
can  possess without having also a continuous symmetry \cite{Ivanov:2012fp} (see also \cite{Darvishi:2019dbh,Darvishi:2021txa}). Among the more predictive potentials are
those where the 3 Higgs fields transform
as a triplet of a non-Abelian discrete symmetry. The highlights of these cases 
are the $A_4$ and $\Delta(27)$ symmetries.

The phenomenology of the $A_4$ invariant potential has been analysed in detail already several years ago \cite{Toorop:2010ex, Toorop:2010kt}.
Meanwhile, while the $\Delta(27)$ symmetry has been often used in particle physics, it was usually done so in a slightly different context (e.g. \cite{Branco:1983tn,deMedeirosVarzielas:2006fc, Ma:2006ip, deMedeirosVarzielas:2011zw,Varzielas:2012nn,Bhattacharyya:2012pi,Ferreira:2012ri,Ma:2013xqa,Nishi:2013jqa,Varzielas:2013sla,Aranda:2013gga,Varzielas:2013eta,Harrison:2014jqa,Ma:2014eka, Fallbacher:2015rea,Abbas:2015zna,Varzielas:2015aua,Bjorkeroth:2015uou,Chen:2015jta,Hernandez:2016eod,CarcamoHernandez:2017owh,deMedeirosVarzielas:2017sdv,Bernal:2017xat,deMedeirosVarzielas:2018vab,CarcamoHernandez:2018djj,Bjorkeroth:2019csz}). Namely, an interesting feature
of a model with such a symmetry is that it exhibits Geometrical CP Violation \cite{Branco:1983tn, deMedeirosVarzielas:2011zw, Varzielas:2012nn,Bhattacharyya:2012pi,Ma:2013xqa, Nishi:2013jqa,Varzielas:2013sla,Varzielas:2013eta,Fallbacher:2015rea},
which means that  there are minima of the potential that violate
CP independently of the parameters of the potential, their form being fixed by the symmetry.

In this paper we therefore aim to fill this gap, by performing a thorough phenomenological analysis of the $\Delta(27)$ invariant potential (with a $\Delta(27)$ triplet of $SU(2)_L$ doublets), which in fact
coincides with  the $\Delta(54)$ invariant potential for the same field content. We refer throughout this work to it as the $\Delta(27)$ potential as we assign fermions also to $\Delta(27)$ representations to study leading order in terms of flavour mixing Yukawa structures and associated flavour violating processes.

We perform an analysis analogous to the study done for the $A_4$ potential in \cite{Toorop:2010ex, Toorop:2010kt}. Namely, we investigate how the parameter space of this potential becomes constrained by $STU$ parameters, perturbative unitarity and, when considering Yukawa structures, the $b \to s \gamma$ process.

In Section \ref{sec:model} we describe the model in detail, covering regions of parameter space, possible minima and relations between the physical parameters (masses etc.) and the parameters of the potential. We also discuss possible leading order extensions of the model to the fermion sector for each minimum. As mentioned before, we extend the model to the fermionic sector by choosing how the fermions transform under the $\Delta(27)$ symmetry, thus determining the structure of the Yukawa couplings. Although realistic fermion mixing requires further symmetry breaking, nevertheless even in this simple implementation we can already discuss some important
features such as the leading aspects of the Higgs mediated flavour changing neutral currents.

In Section \ref{sec:pheno} we present the constraints from $STU$ and unitarity in terms of physical masses of the charged and neutral Higgs bosons.
Finally, we conclude in Section \ref{sec:Conc}.

Appendix~\ref{app:D27} contains the relevant details on the discrete group $\Delta(27)$, while
Appendix~\ref{app:spheno_setup} gives some technical details on \texttt{SPheno} setup and calculation of unitarity constraints.
We also give an example of non-SM Higgs decays pattern in Appendix~\ref{sec:br}.

%% file: tex/model.tex
\section{The $\Delta(27)$ symmetric 3HDM \label{sec:model}}

In this section we discuss the scalar potential, its tree-level minima, Higgs mass matrices as well as the fermion sector of the model.

We consider the most general CP-conserving 3HDM scalar potential, where the Higgs field content is a triplet of $\Delta(27)$ in which  each of the three components of the triplet is an $SU(2)_L$ doublet.
For phenomenological analysis the model is encoded into \sa \cite{Staub:2009bi,Staub:2010jh,Staub:2012pb,Staub:2013tta} version 4.14.5 which, among others, automatically derives aforementioned mass matrices and tadpole equations (minima conditions), as well as interaction vertices.\footnote{The model files are attached to the \texttt{arXiv} version of this work.}

\subsection{Scalar potential}

Using the notation as in \cite{Varzielas:2016zjc, deMedeirosVarzielas:2017glw}, the scalar potential has a contribution that is common to all $\Delta(3n^2)$ ($n\geq 3$) potentials
\begin{align}
V_{0} ( \Phi ) =&
 - \mu^2 \sum_{i, \alpha}   \Phi_{i \alpha}  \Phi^{*i\alpha} + s \sum_{i, \alpha, \beta}  ( \Phi_{i \alpha}  \Phi^{*i\alpha})( \Phi_{i \beta}
\Phi^{*i\beta}) \notag \\
&+ \sum_{i, j, \alpha, \beta}  \left[ r_1 ( \Phi_{i \alpha}  \Phi^{*i\alpha})( \Phi_{j \beta}  \Phi^{*j \beta})
+ r_2 ( \Phi_{i \alpha}  \Phi^{*i \beta})( \Phi_{j \beta}  \Phi^{*j\alpha}) \right],
\label{eq:potV0H}
\end{align}
where the index $i=1,2,3$ refers to the doublet and $\alpha, \beta =1,2$ to the $SU(2)_L$ component of the doublet, either up or down.
$\Delta(3n^2)$ ($n\geq 3$) is a discrete subgroup of the continuous $SU(3)$ group (not to be confused with the $SU(3)_C$ gauge symmetry of the SM), and $V_{0}$ would be invariant under the continuous $SU(3)$ if the coefficient $s$ was set to zero.
The potential that is invariant under $\Delta(27)$  ($\Delta(3n^2)$ with $n=3$) contains $V_{0}$ and in addition the following term 

\begin{align}\label{V27H}
V_{\Delta(27)} ( \Phi ) = V_{\Delta(54)} ( \Phi ) = V_0 ( \Phi )
~+~ \sum_{\alpha, \beta} \left[d \left(
\Phi_{1 \alpha} \Phi_{1 \beta}  \Phi^{*2 \alpha}  \Phi^{*3 \beta} +
\text{cycl.} \right) +
\text{h.c.}\right],
\end{align}
which is not invariant under $A_4$ or $\Delta(3n^2)$ for $n>3$ but still coincides with the
$\Delta(54)$ potential.  Only when we extend the model to the fermionic
sector and choose how the fermions transform under the symmetry, thus determining the
structure of the Yukawa couplings, will it be possible to distinguish the two
types of models.
The parameter $d$ in Eq.~\eqref{V27H} is the only parameter of the potential that can be complex.
Choosing $d$ to be real (as we do in this work) leads to a explicit CP-conserving potential  and it allows for a simple CP transformation under which each doublet
transforms trivially, i.e. each doublet transforms into its complex conjugate. We note that CP can also be conserved with a complex $d$ provided that
its phase is of the form  $\pm i \tfrac{2 \pi}{3}$. In the latter case the CP
transformation is no longer the trivial one and will relate different doublets and their complex conjugates.
The potential (\ref{V27H}) has the interesting feature of having minima with spontaneous geometrical CP violation \cite{Branco:1983tn, deMedeirosVarzielas:2011zw} for real $d$, i.e.,  
for a large region of parameter space there is a CP violating minimum where the phases of the vacuum expectation value (VEV) are fixed to a specific value (in this case, an integer multiple of $2 \pi/3$), 
with this value not depending on  the parameters of the potential (however, for sufficiently large variation of the parameters, one gets into a separate region of parameter 
space, where the minima belong to a different class).

\subsection{The minima of the potential}

After spontaneous gauge symmetry breakdown the Higgs doublets can be
decomposed as
\begin{eqnarray}
\Phi_j = e^{i\alpha_j}   \left( \begin{array}{c}
\phi^+_j \\
\frac{1}{\sqrt{2}}(v_j +  i a_j  + \rho_j + i \eta_j)
\end{array}\right), \qquad j= 1, 2, 3
\end{eqnarray}
with real scalar fields $\rho_j$, $\eta_j$ and $v_j + i a_j$ the (complex) vacuum expectation values (where $v_i$ and $a_i$ are real numbers).

The minima of the potential (\ref{V27H})  have been classified previously \cite{Ivanov:2014doa, deMedeirosVarzielas:2017glw}.
As discussed above, the real  parameter $s$ governs a term that is not $SU(3)$ invariant and that distinguishes directions of VEVs. For $s < 0$, the global minimum favoured is in the $(v,0,0)$ direction, while for $s > 0$ in the $(v,v,v)/\sqrt{3}$ direction. The $d$ parameter (complex in general) governs the phase-dependent term that makes the potential invariant under $\Delta(27)$ (or $\Delta(54)$ to be more precise). If the magnitude of this coefficient is large it disfavours $(v,0,0)$  being the global minimum, as for this VEV $d$ does not contribute to the potential --- a direction like $(v,v,v)/\sqrt{3}$ or $( \omega v,v,v)/\sqrt{3}$ or similar becomes the global minimum when $d$ dominates ($\omega\equiv\exp(2\pi i/3$)). Then, if a CP symmetry is imposed making $d$ a real parameter, the sign of $d$ determines the class of the minima, for positive $d$ the minimum is of the class $(\omega v,v,v)/\sqrt{3}$, with spontaneous geometrical CP violation.

\subsubsection{General VEVs $v_i+i a_i$}

Here we list the extrema conditions for completely general complex VEVs

\begin{align}
\frac{\partial V}{\partial v_1}=&-\mu^2 v_1+sv_1(v_1^2+a_1^2)+(r_1+r_2) v_1 (v_1^2+a_1^2+v_2^2+a_2^2+v_3^2+a_3^2)\\
&+ d [v_1(v_2v_3-a_2a_3)+v_2a_3(a_1+a_2)+a_2v_3(a_1+a_3) +\tfrac{1}{2}v_2(v^2_3-a^2_3)+\tfrac{1}{2}v_3(v_2^2-a_2^2)] , \nonumber \\
\frac{\partial V}{\partial a_1}=&-\mu^2 a_1+sa_1(v_1^2+a_1^2)+(r_1+r_2) a_1 (v_1^2+a_1^2+v_2^2+a_2^2+v_3^2+a_3^2)\\
& + d [a_1(a_2a_3-v_2v_3)+v_2a_3(v_1+v_3)+a_2v_3(v_1+v_2) +\tfrac{1}{2}a_2(a_3^2-v_3^2)+\tfrac{1}{2}a_3(a_2^2-v_2^2)] \nonumber
\end{align}
and the other 4 eqs. obtained by a cyclic permutation of $(123)$.

The above equations are simplified when considering real VEVs. Then we have as equations
\begin{align}
\tfrac{1}{2} d v_1 v_2 (v_1 + v_2 + 2 v_3) + v_3 ( - \mu^2  + s v_3^2 + (r_1 + r_2) (v_1^2 + v_2^2 + v_3^2)) = 0
\end{align}
plus two cyclic permutations. We found several solutions to these equations, which we separate depending on whether their direction does or does not depend directly on the parameters of the potential. Within the solutions we find the (known) minima: $(v,0,0)$ and $(v,v,v)$, as well as solutions of the type $(v,-v,0)$ which we verified analytically can not be minima for any region of parameter space. There are also solutions, whose entries are  complicated functions of the parameters, which we checked numerically are not minima for any of the points in parameter space we sampled. This is in agreement with \cite{Ivanov:2014doa, deMedeirosVarzielas:2017glw}. For complex VEVs we relied on the known solutions $(\omega v, v, v)$ and $(\omega^2 v, v, v)$
(and cyclic permutations) \cite{Ivanov:2014doa, deMedeirosVarzielas:2017glw}. As we are considering cases with CP symmetry of the potential, $(\omega^2 v, v, v)$ is related by the CP symmetry to $(\omega v, v, v)$, so we do not need to consider it separately.

\subsubsection{The $(v,0,0)$ case}
To be a minimum the region of parameter space for $(v,0,0)$ requires
\begin{align}
  \mu^2 = (r_1 + r_2 + s) v^2.
\end{align}
In this case the $\Phi_1$ plays the role of the SM Higgs doublet, while the other two doublets remain VEV-less and couple to gauge bosons
  only through quartic couplings even after the Electroweak Symmetry Breaking (EWSB). Therefore we identify $v$ with the SM VEV, $v \approx 247$ GeV.
The CP-even Higgs mass matrix takes the form 
\begin{align}
m_H^2 =
\begin{pmatrix}
  2 (r_1 + r_2 + s)  & 0 & 0 \\
  0 & - s  & \tfrac{1}{2} d  \\
  0 & \tfrac{1}{2} d  & - s 
\end{pmatrix} v^2
\label{15}
\end{align}
with eigenvalues
$m_h^2=2 (r_1 + r_2 + s) v^2$, $m^2_{H_1}=\tfrac{1}{2}(d - 2 s) v^2$, $m^2_{H_2}=-\tfrac{1}{2} (d +2 s) v^2$, where $m_{H_1}$ and $m_{H_2}$ are not mass ordered, and $h$ is to be identified with the SM-like Higgs.

The CP-odd Higgs mass matrix is given by 
\begin{align}
m_A^2 =
\begin{pmatrix}
  0 & 0 & 0 \\
  0 & - s  & -\tfrac{1}{2} d  \\
  0 & -\tfrac{1}{2} d  & - s 
\end{pmatrix}  v^2 + \text{gauge dependent terms}
\end{align}
with eigenvalues  $m_{A_1}^2 =  \tfrac{1}{2} (d - 2 s)v^2$, $m_{A_2}^2 = -\tfrac{1}{2} (d + 2 s)v^2$ and a Goldstone boson $G^0$.
The $m_{H^\pm}^2$ matrix is diagonal, with $m_{H^\pm_{1}}^2 =m_{H^\pm_{2}}^2 = -(r_2 + s) v^2$ and a Goldstone boson $G^\pm$.

The Lagrangian parameters can be expressed in terms of physical masses of Higgses as
\begin{align}
\label{eq:pot_pars_100_1}
  s = & -\frac{1}{2 v^2} (m_{H_1}^2 + m_{H_2}^2),\\
  r_1 = & \mathbin{\hphantom{-}}\frac{1}{2 v^2} (m_h^2 + 2 m_{H^\pm}^2), \\
  r_2 = & \mathbin{\hphantom{-}}\frac{1}{2 v^2} (m_{H_1}^2 + m_{H_2}^2 - 2 m_{H^\pm}^2),\\
\label{eq:pot_pars_100_2}
  d = & \mathbin{\hphantom{-}}\frac{1}{v^2} (m_{H_1}^2 - m_{H_2}^2). 
\end{align}
Here $H^\pm$ is a simplified notation for either $H^\pm_1$ or $H^\pm_2$, which are degenerate in mass. Likewise 
$A_i$ are pairwise mass degenerate with $H_i$.

\subsubsection{The  $(v,v,v)$ case}
To be a minimum the region of parameter space for $(v,v,v)$ requires
\begin{align}
\mu^2 = (2 d + 3 r_1 + 3 r_2 + s) v^2.
\end{align}
The CP-even Higgs mass matrix takes the form $(r_{12}\equiv r_1+r_2)$
\begin{align}
m_H^2 = \frac{1}{2}
\begin{pmatrix}
4 (r_{12}+s)-2 d & 5 d+4 r_{12} &
   5 d+4 r_{12} \\
 5 d+4 r_{12} & 4 (r_{12}+s)-2d &
   5 d+4 r_{12} \\
 5 d+4 r_{12}& 5 d+4
   r_{12}& 4 (r_{12}+s)-2d
\end{pmatrix} v^2
\end{align}
with eigenvalues 
$m_h^2=2 (2 d + 3 r_{12} +  s) v^2$
(to be identified with the SM-like Higgs boson mass) and  $m_H^2=1/2 (-7 d +   4 s) v^2$ (the mass of the pair of degenerate Higgs bosons $H_1$ and $H_2$).

The CP-odd mass matrix is given by
\begin{align}
m_A^2 = \frac{3}{2}
\begin{pmatrix}
 -2 & 1 & 1 \\
 1 & -2 & 1 \\
 1 & 1 & -2 \\
\end{pmatrix} d v^2 + \text{gauge dependent terms} .
\end{align}
The two physical CP-odd Higgs bosons $A_1$ and $A_2$ are mass degenerate with a common mass $m_A^2=-9dv^2/2$.
Similarly the charged Higgs bosons mass matrix is given by
\begin{align}
m_{H^\pm}^2 =
\begin{pmatrix}
 -2 & 1 & 1 \\
 1 & -2 & 1 \\
 1 & 1 & -2 \\
\end{pmatrix} (r_2 + d) v^2 + \text{gauge dependent terms},
\end{align}
where $H_1^\pm$ and $H_2^\pm $ are also mass degenerate, with a common mass $m_{H^\pm}^2 = - 3(d + r_2) v^2$.

The Lagrangian parameters can  be expressed in terms of physical masses of Higgses as
\begin{align}
\label{eq:pot_pars_111_1}
  s = & \mathbin{\hphantom{-}}\frac{1}{18 v^2} (9 m_H^2 - 7 m_A^2),\\
  r_1 = & \mathbin{\hphantom{-}}\frac{1}{18 v^2} (m_A^2 + 6 m_{H^\pm}^2 + 3 m_h^2 - 3 m_H^2),\\
  r_2 = & \mathbin{\hphantom{-}}\frac{1}{9 v^2} (2 m_A^2 - 3 m_{H^\pm}^2),\\
\label{eq:pot_pars_111_2}
  d = & -\frac{2}{9 v^2} m_A^2.
\end{align}

\subsubsection{The $(\omega v, v,v)$ case \label{sec:omega}}

To be a minimum the region of parameter space for $(\omega v,v,v)$ requires\footnote{Because we are taking $d$ to be real, the potential is invariant under the trivial CP symmetry (as discussed above), so the $(\omega^2 v, v,v)$ case (which is in general distinct from this case) collapses into the same orbit as $(\omega v, v,v)$ and we do not discuss it here separately.}
\begin{align}
      \mu^2 =& (-d + 3 r_{12} + s) v^2.
\end{align}
Because in this vacuum CP is spontaneously broken, this time all Higgs bosons mix and the mass matrix for the neutral ones takes the form
\begin{align}
m_H^2 =
\begin{pmatrix}
m_{HH}^2 & m_{HA}^2 \\
(m_{HA}^2)^T & m_{AA}^2
\end{pmatrix} v^2 + \text{gauge dependent terms},
\end{align}
where the 3x3 submatrices are as follows
\begin{align}
m_{HH}^2 =&
\begin{pmatrix}
 \frac{1}{2} \left(4 d+r_{12}+s\right) & d-r_{12} & d-r_{12} \\
 d-r_{12} & \frac{1}{2} \left(d+4 \left(r_{12}+s\right)\right) & \frac{1}{4} \left(8 r_{12}-5 d\right) \\
 d-r_{12} & \frac{1}{4} \left(8 r_{12}-5 d\right) & \frac{1}{2} \left(d+4 \left(r_{12}+s\right)\right)
\end{pmatrix}, \\
m_{HA}^2 =& \frac{\sqrt{3}}{2} 
\begin{pmatrix}
 - \left(r_{12}+s\right) & - d+2 r_{12} & - d +2 r_{12} \\
 d &  d & - d/2 \\
 d & -d/2 & d
\end{pmatrix}, \\
m_{AA}^2 =& \frac{3}{2}
\begin{pmatrix}
r_{12}+s & 0 & 0 \\
 0 & d & -d/2 \\
 0 & - d/2 & d
\end{pmatrix}
\end{align}
with the eigenvalues
\begin{align}
m_{h}^2& = 2 v^2 \left(-d+3 r_1+3 r_2+s\right),\\
m_{H_1}^2& = m_{H_2}^2 = \frac{1}{2} \left(4 d+2 s - \sqrt{7 d^2-2 d s+4 s^2}\right)v^2 ,\\
m_{H_3}^2 &= m_{H_4}^2 = \frac{1}{2} \left(4 d+2 s + \sqrt{7 d^2-2 d s+4 s^2}\right)v^2 .
\end{align}
We assume that the single mass-nondegenerate state  $h$ is identified with the SM-like Higgs.

The mass matrix for the charged Higgs bosons  reads
\begin{align}
m_{H^- H^+}^2 =  ( d-2 r_2 )
\begin{pmatrix}
1 & -\tfrac{\omega}{2} & -\tfrac{\omega}{2} \\
-\frac{\omega^*}{2} \ &  1 & -\tfrac{1}{2} \\
-\tfrac{\omega^*}{2} & -\tfrac{1}{2} & 1
\end{pmatrix}
v^2 + \text{gauge dependent terms} .
\end{align}
Physical charged states are mass degenerate, with mass
$m_{H^\pm_1}^2 = m_{H^\pm_2}^2 = \frac{3}{2} (d-2r_2)v^2$
.

Because of the quadratic nature of expressions for masses $m_{H_i}^2$ there are two combinations of potential parameters that give the same 4 masses $m_h$, $m_{H_1}$, $m_{H_3}$ and $m_{H^\pm}$, namely
\begin{align}
\label{eq:pot_pars_o11_1}
 s =& \frac{1}{18 v^2} \left(3 m_{H_1}^2 + 3 m_{H_3}^2 \mp 2 \Omega \right),\\
 r_1 =& \frac{1}{18 v^2} \left(6 m_h^2 - 3 m_{H_1}^2 - 3 m_{H_3}^2  + 12 m_{H^\pm}^ 2 \pm \Omega \right),\\
 r_2 =& \frac{1}{18 v^2} \left(3 m_{H_1}^2 + 3 m_{H_3}^2  - 12 m_{H^\pm}^2\pm \Omega  \right),\\
\label{eq:pot_pars_o11_2}
 d =& \frac{1}{18 v^2} \left(3 m_{H_1}^2 + 3 m_{H_3}^2 \pm \Omega \right),
\end{align}
where $m_{H_3}^2 > 3 m_{H_1}^2$ in order for $\Omega\equiv \sqrt{9m_{H_1}^4 - 30 m_{H_1}^2 m_{H_3}^2+ 9 m_{H_3}^4 }$ to be real.

\subsection{Yukawa couplings}

Having discussed the scalar potential and its vacua, we now proceed with description of the Yukawa interactions.

The general form of the quark Yukawa couplings in models with three Higgs doublets is
\begin{align}
L_Y =&  -\overline{Q_L} \ \Gamma_1 \Phi_1 d_R - \overline{Q_L}\
\Gamma_2 \Phi_2 d_R - \overline{Q_L} \ \Gamma_3 \Phi_3 d_R  \nonumber \\
& - \overline{Q_L} \ \Delta_1 \tilde{\Phi}_1 u_R -
\overline{Q_L} \ \Delta_2 \tilde{\Phi}_2 u_R \
- \overline{Q_L} \ \Delta_3 \tilde{\Phi}_3 u_R \ + \mbox{h.c.},
\label{1e2}
\end{align}
where $\Gamma_i$ and $\Delta_i$ denote the Yukawa couplings of the
left-handed (LH) quark doublets $Q_L$ to the Higgs doublets $\Phi_j$ and, respectively, right-handed (RH) quarks $d_R$ or $u_R$.
After spontaneous symmetry breaking quark mass matrices are generated with the form
\begin{eqnarray}
M_d &=& \frac{1}{\sqrt{2}} ( v_1  e^{i \alpha_1} \Gamma_1  +
                           v_2 e^{i \alpha_2} \Gamma_2 + v_3 e^{i \alpha_3} \Gamma_3 ), \nonumber \\
M_u &=& \frac{1}{\sqrt{2}} ( v_1  e^{- i \alpha_1} \Delta_1 +
                           v_2 e^{-i \alpha_2} \Delta_2 +  v_3 e^{-i \alpha_3} \Delta_3 ).
\label{mmmm}
\end{eqnarray}
The matrices $M_d, M_u$ are then diagonalised by the usual
bi-unitary transformations
\begin{eqnarray}
U^\dagger_{dL} M_d U_{dR} = D_d \equiv {\mbox diag}\ (m_d, m_s, m_b),
\label{umu}\\
U^\dagger_{uL} M_u U_{uR} = D_u \equiv {\mbox diag}\ (m_u, m_c, m_t).
\label{uct}
\end{eqnarray}

Before we can continue with the phenomenological analysis we need to assign the fermions to $\Delta(27)$ representations to obtain semi-realistic Yukawa couplings.
The RH quarks we choose to be $\Delta(27)$ triplets because, as we present in more detail for each of the VEVs, this choice allows us to make semi-realistic $\Delta(27)$ invariants for both up and down quark sectors.
The 3 LH quark doublets will be distinct $\Delta(27)$ singlets but the choice of representations depends on the VEV, in order to have Yukawa structures that are realistic at leading order - we are able to obtain distinct masses for each generation but with a CKM matrix that is the unit matrix, which is a consequence of breaking $\Delta(27)$ with just one triplet scalar that leaves too much residual unbroken flavour symmetry. While such mixing angles are clearly not viable, we consider this to be a good leading-order approximation. We consider these Yukawa structures as toy models to obtain constraints on the scalar sector. More realistic mixing angles can be obtained by adding further sources of $\Delta(27)$ breaking \cite{Bhattacharyya:2012pi,Varzielas:2013sla,Varzielas:2013eta}.

In the following subsections we present Yukawa couplings as $Y_u$, $Y_d$ and $Y_e$ matrices, and later show the respective $\Gamma$, $\Delta$ matrices that couple to each of the $\Phi_i$ fields.

\subsubsection{The  $(v,0,0)$ case}

For the simplest VEV, the choice of singlets is one trivial ($1_{00}$) and two non-trivial  $1_{0i}$ ($i=1,2$). This choice corresponds to having the three singlets that do not transform under the generator $c$ of $\Delta(27)$ (see Appendix \ref{app:D27}), such that we have the desired composition rules from the product of the (anti-)triplet RH quark (either $d_R$ or
$u_R$) and the (anti-)triplet Higgs (either $\Phi_j$ for the down sector or its conjugate for the up sector).

In particular, with $(d_R)_i$ ($i=1\ldots3$, the generation index) transforming as an anti-triplet $\bar{3}$, the $\Delta(27)$ invariant terms are then of the type $\bar{Q}_i (\Phi d_R)_{0j}$, whereas $(u_R)_i$ should transform as a triplet $3$ to construct similarly the invariants with $\tilde{\Phi}$ which transforms as an anti-triplet $\bar{3}$, i.e. $\bar{Q}_i (u_R \tilde{\Phi})_{0j}$.

For $\bar{Q}_1, \bar{Q}_2, \bar{Q}_3$ belonging to, respectively, $1_{00}, 1_{02}$ and $1_{01}$ the corresponding Yukawa terms are
\begin{align}
L_Y^d=\bar{Q}Y_d d_R \equiv&y_1^d \bar{Q}_1 \left( \Phi_1 (d_R)_1 + \Phi_2 (d_R)_2 + \Phi_3 (d_R)_3 \right)_{00} + \nonumber \\
&y_2^d \bar{Q}_2 \left( \Phi_3 (d_R)_1 + \Phi_1 (d_R)_2 + \Phi_2 (d_R)_3 \right)_{01} + \nonumber \\
&y_3^d \bar{Q}_3 \left( \Phi_2 (d_R)_1 + \Phi_3 (d_R)_2 + \Phi_1 (d_R)_3 \right)_{02}.
\end{align}
These expressions follow from the specific choice of non-trivial singlets made for the LH quarks and the composition rules of $\Delta(27)$ listed in Appendix \ref{app:D27}.

The Yukawa terms above correspond to Yukawa matrices of the form (in LR convention)
\begin{align}
Y_d=
\begin{pmatrix}
y_1^d \Phi_1 & y_1^d \Phi_2 & y_1^d \Phi_3 \\
y_2^d \Phi_3 & y_2^d \Phi_1 & y_2^d \Phi_2 \\
y_3^d \Phi_2 & y_3^d \Phi_3 & y_3^d \Phi_1
\end{pmatrix}.
\end{align}

The mass matrices are now easy to read of when $\Phi$ acquires its VEV, breaking $SU(2)_L$ and $\Delta(27)$. In this subsection we are dealing with the $(v,0,0)$ VEV, which means that  we are already in the Higgs basis. The down quark mass matrix then looks as follows
\begin{align}
M_d=
\begin{pmatrix}
y_1^d v & 0 & 0 \\
0 & y_2^d v & 0 \\
0 & 0 & y_3^d v
\end{pmatrix}.
\end{align}

Because the LH quarks are $\Delta(27)$ singlets, it is easy to adapt the above assignments to the up quark sector giving
\begin{align}
L_Y^u=&y_1^u \bar{Q}_1 \left( \tilde{\Phi}_1 (u_R)_1 + \tilde{\Phi}_2 (u_R)_2 + \tilde{\Phi}_3 (u_R)_3 \right)_{00} + \nonumber \\
&y_2^u \bar{Q}_2 \left( \tilde{\Phi}_2 (u_R)_1 + \tilde{\Phi}_3 (u_R)_2 + \tilde{\Phi}_1 (u_R)_3 \right)_{01} + \nonumber \\
&y_3^u \bar{Q}_3 \left( \tilde{\Phi}_3 (u_R)_1 + \tilde{\Phi}_1 (u_R)_2 + \tilde{\Phi}_2 (u_R)_3 \right)_{02}.
\end{align}
Note the change compared to the down sector. We have to construct the same singlets of $\Delta(27)$ from $3 \times \bar{3}$ (see Appendix \ref{app:D27}), but in the down sector the scalar $\Phi$ is in the $3$ and the RH fermion is in the $\bar{3}$, whereas in the up sector the scalar $\tilde{\Phi}$ is in the $\bar{3}$ and it is the RH fermion that transforms as $3$. Thus
\begin{align}
Y_u=
\begin{pmatrix}
y_1^u \tilde{\Phi}_1 & y_1^u \tilde{\Phi}_2 & y_1^u \tilde{\Phi}_3 \\
y_2^u \tilde{\Phi}_2 & y_2^u \tilde{\Phi}_3 & y_2^u \tilde{\Phi}_1 \\
y_3^u \tilde{\Phi}_3 & y_3^u \tilde{\Phi}_1 & y_3^u \tilde{\Phi}_2
\end{pmatrix}.
\end{align}
For the $(v,0,0)$ VEV we therefore have
\begin{align}
M_u=
\begin{pmatrix}
y_1^u v & 0 & 0 \\
0 & 0 & y_2^u v \\
0 & y_3^u v & 0
\end{pmatrix},
\end{align}
which gives
\begin{align}
M_u M_u^\dagger=
\begin{pmatrix}
|y_1^u v|^2 & 0 & 0 \\
0 & |y_2^u v|^2 & 0 \\
0 & 0 & |y_3^u v|^2
\end{pmatrix}.
\end{align}

Given that both $M_u M_u^\dagger$ and $M_d M_d^\dagger$ are diagonal, the CKM matrix in this limit is the identity matrix.
For this phenomenological study we consider this as a reasonable leading order approximation.

The extension to the charged lepton sector is likewise simple, where we choose to take the $SU(2)_L$ doublet $L$ as a triplet so that the RH charged leptons are singlets. The charged lepton masses can be easily obtained in this basis, e.g. with $e^c_1, e^c_2, e^c_3$ respectively transforming as $\Delta(27)$ singlets $1_{00}, 1_{02}, 1_{01}$
\begin{align}
L_Y^e=
&y_1^e \left( \bar{L}_1 \Phi_1 + \bar{L}_2 \Phi_2 + \bar{L}_3 \Phi_3 \right)_{00} e^c_1 + \nonumber \\
&y_2^e \left( \bar{L}_1 \Phi_3 + \bar{L}_2 \Phi_1 + \bar{L}_3 \Phi_2 \right)_{01} e^c_2 + \nonumber \\
&y_3^e \left( \bar{L}_1 \Phi_2 + \bar{L}_2 \Phi_3 + \bar{L}_3 \Phi_1 \right)_{02} e^c_3,
\end{align}
corresponding to Yukawa matrices of the form (LR convention)
\begin{align}
Y_e=
\begin{pmatrix}
y_1^e \Phi_1 & y_2^e \Phi_3 & y_3^e \Phi_2 \\
y_1^e \Phi_2 & y_2^e \Phi_1 & y_3^e \Phi_3 \\
y_1^e \Phi_3 & y_2^e \Phi_2 & y_3^e \Phi_1
\end{pmatrix},
\end{align}
leading again to a diagonal matrix for the $(v,0,0)$ VEV
\begin{align}
M_e=
\begin{pmatrix}
y_1^e v & 0 & 0 \\
0 & y_2^e v & 0 \\
0 & 0 & y_3^e v
\end{pmatrix}.
\end{align}

This alternative choice of $\bar{L} \sim 3_{02}$ allows for interesting possibilities for obtaining large leptonic mixing, see e.g. \cite{Varzielas:2013eta}, but as this depends on the mechanism that gives neutrinos their masses (the type of seesaw for example), it is beyond the scope of this paper.

\subsubsection{The $(v,v,v)$ case }

For this VEV we use singlets transforming only under the generator of $\Delta(27)$ which we refer to as the $c$ generator (see Appendix~\ref{app:D27}).

With $d_R$ transforming as a $\bar{3}$, the $\Delta(27)$ invariant terms are then of the type $\bar{Q}_i (H (d_R))_{j0}$, whereas $u_R$ should transform as a $3$ to construct similarly the invariants with $\tilde{\Phi}$ which transforms as a $\bar{3}$, i.e. $\bar{Q}_i ((u_R) \tilde{\Phi})_{j0}$.

For $\bar{Q}_1, \bar{Q}_2, \bar{Q}_3$ chosen as, respectively, $1_{00}, 1_{20}, 1_{10}$ (note these are not the same choices as above), the corresponding Yukawa terms expanded are
\begin{align}
L_Y^d=
&y_1^d \bar{Q}_1 \left( \Phi_1 (d_R)_1 + \Phi_2 (d_R)_2 + \Phi_3 (d_R)_3 \right)_{00} + \nonumber \\
&y_2^d \bar{Q}_2 \left( \Phi_1 (d_R)_1 + \omega^2 \Phi_2 (d_R)_2 + \omega \Phi_3 (d_R)_3 \right)_{10} + \nonumber \\
&y_3^d \bar{Q}_3 \left( \Phi_1 (d_R)_1 + \omega \Phi_2 (d_R)_2 + \omega^2 \Phi_3 (d_R)_3 \right)_{20}
\end{align}
corresponding to Yukawa matrices of the form (LR convention)
\begin{align}
\label{eq:yuk111}
Y_d=
\begin{pmatrix}
y_1^d \Phi_1 & y_1^d \Phi_2 & y_1^d \Phi_3 \\
y_2^d \Phi_1 & \omega^2 y_2^d \Phi_2 & \omega y_2^d \Phi_3 \\
y_3^d \Phi_1 & \omega y_3^d \Phi_2 & \omega^2 y_3^d \Phi_3
\end{pmatrix}.
\end{align}

The mass matrices are now easy to construct when $\Phi$ acquires its VEV, breaking $SU(2)$ and $\Delta(27)$. With the $(v,v,v)$ VEV, we are not in the Higgs basis. The down mass matrix looks like
\begin{align}
M_d=
\begin{pmatrix}
y_1^d v & y_1^d v & y_1^d v \\
y_2^d v & \omega^2 y_2^d v & \omega y_2^d v\\
y_3^d v & \omega y_3^d v & \omega^2 y_3^d v,
\end{pmatrix},
\end{align}
which also in this case is not diagonal, but gives
\begin{align}
M_d M_d^\dagger=
\begin{pmatrix}
|y_1^d v|^2 & 0 & 0 \\
0 & |y_2^d v|^2 & 0 \\
0 & 0 & |y_3^d v|^2
\end{pmatrix}.
\end{align}

Because the LH quarks are singlets, it is easy to adapt the above assignments to the up-quark sector. In this case the change of roles of triplet and anti-triplets does not change the type of invariant, and we have
\begin{align}
L_Y^u=
&y_1^u \bar{Q}_1 \left( \tilde{\Phi}_1 (u_R)_1 + \tilde{\Phi}_2 (u_R)_2 + \tilde{\Phi}_3 (u_R)_3 \right)_{00} + \nonumber \\
&y_2^u \bar{Q}_2 \left( \tilde{\Phi}_1 (u_R)_1 + \omega^2 \tilde{\Phi}_2 (u_R)_2 + \omega \tilde{\Phi}_3 (u_R)_3 \right)_{10} + \nonumber \\
&y_3^u \bar{Q}_3 \left( \tilde{\Phi}_1 (u_R)_1 + \omega \tilde{\Phi}_2 (u_R)_2 + \omega^2 \tilde{\Phi}_3 (u_R)_3 \right)_{20}
\end{align}
and
\begin{align}
Y_u=
\begin{pmatrix}
y_1^u \tilde{\Phi}_1 & y_1^u \tilde{\Phi}_2 & y_1^u \tilde{\Phi}_3 \\
y_2^u \tilde{\Phi}_1 & \omega^2 y_2^u \tilde{\Phi}_2 & \omega y_2^u \tilde{\Phi}_3 \\
y_3^u \tilde{\Phi}_1 & \omega y_3^u \tilde{\Phi}_2 & \omega^2 y_3^u \tilde{\Phi}_3
\end{pmatrix}.
\end{align}
For the same VEV as before we have the matrix
\begin{align}
M_u=
\begin{pmatrix}
y_1^u v & y_1^u v & y_1^u v \\
y_2^u v & \omega^2 y_2^u v & \omega y_2^u v \\
y_3^u v & \omega y_3^u v & \omega^2 y_3^u v
\end{pmatrix},
\end{align}
which gives
\begin{align}
M_u M_u^\dagger=
\begin{pmatrix}
|y_1^u v|^2 & 0 & 0 \\
0 & |y_2^u v|^2 & 0 \\
0 & 0 & |y_3^u v|^2
\end{pmatrix}.
\end{align}

For the charged leptons and this VEV choice we can not mimic a transposed down sector as we had done for $(v,0,0)$. We instead assign similarly the $e^c$ as an anti-triplet and assign the $\bar{L}_1, \bar{L}_2, \bar{L}_3$, respectively, as $1_{00}, 1_{20}, 1_{10}$.
With this choice we get
\begin{align}
L_Y^e=
&y_1^d \bar{L}_1 \left( \Phi_1 e^c_1 + \Phi_2 e^c_2 + \Phi_3 e^c_3 \right)_{00} + \nonumber\\
&y_2^d \bar{L}_2 \left( \Phi_1 e^c_1 + \omega^2 \Phi_2 e^c_2 + \omega \Phi_3 e^c_3 \right)_{10} + \nonumber\\
&y_3^d \bar{L}_3 \left( \Phi_1 e^c_1 + \omega \Phi_2 e^c_2 + \omega^2 \Phi_3 e^c_3 \right)_{20}
\end{align}
corresponding to Yukawa matrices of the form (LR convention)
\begin{align}
Y_e=
\begin{pmatrix}
y_1^e \Phi_1 & y_1^e \Phi_2 & y_1^e \Phi_3 \\
y_2^e \Phi_1 & \omega^2 y_2^e \Phi_2 & \omega y_2^e \Phi_3 \\
y_3^e \Phi_1 & \omega y_3^e \Phi_2 & \omega^2 y_3^e \Phi_3
\end{pmatrix}
\end{align}
leading to a non-diagonal matrix $M_e$ for the $(v,v,v)$ VEV, but to a diagonal $M_e M_e^\dagger$ combination.

\subsubsection{The  $(\omega v,v,v)$ case }

For $\bar{Q}_1, \bar{Q}_2, \bar{Q}_3$ chosen as, respectively, $1_{00}, 1_{02}, 1_{01}$ (i.e. the same choices as taken above for the $(v,0,0)$) the corresponding Yukawa matrices  are the same in terms of $\Phi_i$
\begin{align}
Y_d=
\begin{pmatrix}
\label{eq:Yd}
y_1^d \Phi_1 & y_1^d \Phi_2 & y_1^d \Phi_3 \\
y_2^d \Phi_3 & y_2^d \Phi_1 & y_2^d \Phi_2 \\
y_3^d \Phi_2 & y_3^d \Phi_3 & y_3^d \Phi_1
\end{pmatrix},
\end{align}

\begin{align}
Y_u=
\begin{pmatrix}
y_1^u \tilde{\Phi}_1 & y_1^u \tilde{\Phi}_2 & y_1^u \tilde{\Phi}_3 \\
y_2^u \tilde{\Phi}_2 & y_2^u \tilde{\Phi}_3 & y_2^u \tilde{\Phi}_1 \\
y_3^u \tilde{\Phi}_3 & y_3^u \tilde{\Phi}_1 & y_3^u \tilde{\Phi}_2
\end{pmatrix}.
\end{align}
Replacing field vector $\Phi_i$ with the VEV we get
\begin{align}
M_d=
\begin{pmatrix}
y_1^d \omega v & y_1^d v & y_1^d v \\
y_2^d v & y_2^d \omega v & y_2^d v \\
y_3^d v & y_3^d v & y_3^d \omega v
\end{pmatrix},
\end{align}
\begin{align}
M_u=
\begin{pmatrix}
y_1^u \omega^2 v & y_1^u v & y_1^u v \\
y_2^u v & y_2^u v & y_2^u \omega^2 v \\
y_3^u v & y_3^u \omega^2 v & y_3^u v
\end{pmatrix}
\end{align}
giving (as in the previous cases) the diagonal products
\begin{align}
M_d M_d^\dagger=
\begin{pmatrix}
|y_1^d v|^2 & 0 & 0 \\
0 & |y_2^d v|^2 & 0 \\
0 & 0 & |y_3^d v|^2
\end{pmatrix},
\end{align}
\begin{align}
M_u M_u^\dagger=
\begin{pmatrix}
|y_1^u v|^2 & 0 & 0 \\
0 & |y_2^u v|^2 & 0 \\
0 & 0 & |y_3^u v|^2
\end{pmatrix}.
\end{align}

For charged leptons and for this VEV we choose to assign the $e^c$ as a $\bar{3}$, and the three $\bar{L}_1, \bar{L}_2, \bar{L}_3$, respectively, as singlets $1_{00}, 1_{02}, 1_{01}$. We therefore obtain exactly the same structures as for the down quarks
\begin{align}
Y_e=
\begin{pmatrix}
y_1^e \Phi_1 & y_1^e \Phi_2 & y_1^e \Phi_3 \\
y_2^e \Phi_3 & y_2^e \Phi_1 & y_2^e \Phi_2 \\
y_3^e \Phi_2 & y_3^e \Phi_3 & y_3^e \Phi_1
\end{pmatrix},
\end{align}
\begin{align}
M_e=
\begin{pmatrix}
y_1^e \omega v & y_1^e v & y_1^e v \\
y_2^e v & y_2^e \omega v & y_2^e v \\
y_3^e v & y_3^e v & y_3^e \omega v
\end{pmatrix},
\end{align}
\begin{align}
M_e M_e^\dagger=
\begin{pmatrix}
|y_1^e v|^2 & 0 & 0 \\
0 & |y_2^e v|^2 & 0 \\
0 & 0 & |y_3^e v|^2
\end{pmatrix}.
\end{align}

\subsection{Flavour Changing Neutral Currents \label{subsec:FCNC}}
In order to see the structure of Flavour Changing Neutral Currents (FCNCs) in this model, let us make the following transformation among the $\Phi_j$ fields
\begin{equation}
\Phi^\prime = O \cdot K \ \Phi,
\end{equation}
with the matrices $O$ and $K$ given by
\begin{eqnarray}
O  = \left(\begin{array}{ccc}
\frac{v_1}{v}  &\frac{v_2}{v}   & \frac{v_3}{v}  \\
\frac{v_2}{v^\prime}  & -\frac{v_1}{v^\prime} & 0  \\
\frac{v_1}{v^{\prime \prime}} & \frac{v_2}{v^{\prime \prime}} &
\frac{ - (v^2_1 + v^2_2)/v_3}{v^{\prime \prime}}
\end{array}\right), \qquad
K=  \left(\begin{array}{ccc}
 e^{- i\alpha_1} & 0 & 0 \\
0 &  e^{- i\alpha_2} & 0 \\
0 & 0 &  e^{-i\alpha_3} \end{array}\right),
\end{eqnarray}
where $v=\sqrt{v^2_1 + v^2_2 + v^2_3} $, $v^\prime = \sqrt{v^2_1 + v^2_2}$
and $v^{\prime \prime} = \sqrt{v^2_1 + v^2_2 + (v^2_1 + v^2_2)^2 / v^2_3}$.
The new components of the $\Phi^\prime$ doublets are the primed scalar fields,
together with $G^0$ and $G^+$
\begin{eqnarray}
\left( \begin{array}{c}
h^\prime \\
R \\
R^\prime
\end{array}\right) = O\ \left( \begin{array}{c}
\rho_1 \\
\rho_2\\
\rho_3
\end{array}\right), \qquad \left( \begin{array}{c}
G^0 \\
I \\
I^\prime
\end{array}\right) = O \ \left( \begin{array}{c}
\eta_1 \\
\eta_2\\
\eta_3
\end{array}\right),
\qquad \left( \begin{array}{c}
G^+ \\
H^{\prime +}_1\\
H^{\prime +}_2
\end{array}\right) = O \ \left( \begin{array}{c}
\phi^+_1 \\
\phi^+_2\\
\phi^+_3
\end{array} \right).
\end{eqnarray}
This transformation singles out
$h^\prime$ as well as  the neutral pseudo-Goldstone boson $G^0$ and
the charged Goldstone boson $G^+$. The scalar field $h^\prime$ has couplings to the
quarks which are proportional to the mass matrices and it is the only scalar field 
in this basis with triple couplings to a pair of gauge bosons. 
The other scalar fields only couple to a pair of
gauge bosons through quartic couplings. As a result $h^\prime$ could be identified as
the SM like Higgs boson if it were already a mass eigenstate.
This fact results from the choice of the first
row of the matrix $O$ \cite{Donoghue:1978cj,Georgi:1978ri},
the choice of the other two rows is free provided that
they respect the orthogonality relations.
Therefore a transformation of this form
may lead to many different scalar bases. It should be noticed that what characterises
the rotation by the matrix $O \cdot K$ is the fact
that in this new scalar basis only the first doublet acquires a VEV, $v$, different from zero. In these bases, the vacuum is of the form $(v, 0, 0)$, with $v$ real.

In general three Higgs doublet models,
$h^\prime$  obtained after this rotation,
is not yet a mass eigenstate. In the CP-conserving case, in general, the physical
neutral scalars are obtained, after
further mixing among $h^\prime$, $R$ and $R^\prime$, as well as mixing
between $I$ and $I^\prime$.
In the CP violating case all five neutral fields may mix among themselves.
In this scalar basis, after the $O$ rotation,
FCNCs arise
from the couplings to the remaining four neutral Higgs fields.
The structure of the Higgs mediated FCNCs and the charged Higgs couplings to the quark mass eigenstates in models with three Higgs doublets are given by \cite{Botella:2009pq}

\begin{eqnarray}
L_Y  &=& - \frac{\sqrt{2} H^{\prime +}_1}{v^\prime} \bar{u}^p \left(
V {\cal{N}}_d \gamma_R + {\cal{N}}^\dagger_u \ V \gamma_L \right) d^p
- \frac{\sqrt{2} H^{\prime +}_2}{v^{\prime \prime}} \bar{u}^p \left(
V {\cal{N}}^\prime_d \gamma_R + {\cal{N}}^{\prime \dagger}_u \ V \gamma_L \right) d^p
\nonumber \\
& - & \frac{h^{\prime}}{v} \left(  \bar{d_L}^p D_d \ d_R^p + \bar{u_L}^p D_u u_R^p  \right)
- \bar{d_L}^p D_d \ d_R^p - \bar{u_L}^p D_u u_R^p
\nonumber \\
& - &  \bar{d_L}^p \frac{1}{v^\prime} \  {\cal{N}}_d (R+iI) d_R^p -
 \bar{u_L}^p \frac{1}{v\prime }  {\cal{N}}_u (R-iI) u_R^p - \\
& - &  \bar{d_L}^p \frac{1}{v^{\prime \prime}} \
 {\cal{N}}^\prime_d (R^\prime +iI^\prime ) d_R^p -
  \bar{u_L}^p \frac{1}{v^{\prime \prime}} \
 {\cal{N}}^\prime_u (R^\prime -iI^\prime ) u_R^p
 + \mbox{h.c.},
\nonumber
\end{eqnarray}
where we denote the physical quark mass eigenstates with a superscript $p$ (e.g. $d_R^p$), and
with
\begin{eqnarray}
 {\cal{N}}_d & =&
\frac{1}{\sqrt{2}} U^\dagger_{dL}\
( v_2 e^{i \alpha_1}  \Gamma_1  -
                           v_1 e^{i \alpha_2} \Gamma_2 )\  U_{dR}, \label{Nd_general} \\
 {\cal{N}}_u &= &\frac{1}{\sqrt{2}} U^\dagger_{uL} ( v_2  e^{-i \alpha_1} \Delta_1  -
                           v_1 e^{-i \alpha_2} \Delta_2 )\  U_{uR}, \label{Nu_general}\\
 {\cal{N}}^\prime_d &=&  \frac{1}{\sqrt{2}} U^\dagger_{dL}\
( v_1  e^{i \alpha_1} \Gamma_1   +
                           v_2 e^{i \alpha_2} \Gamma_2
+ x  e^{i \alpha_3} \Gamma_3  )\  U_{dR}, \label{Ndp_general} \\
 {\cal{N}}^\prime_u &=&  \frac{1}{\sqrt{2}} U^\dagger_{uL}\
( v_1 e^{-i \alpha_1} \Delta_1  +
                           v_2 e^{-i \alpha_2} \Delta_2
+ x  e^{-i \alpha_3} \Delta_3  )\  U_{uR},
\label{Nup_general}
\end{eqnarray}
where $x \equiv - (v^2_1 + v^2_2)/v_3$.

For completeness, let us consider $ {\cal{N}}_d $ and $ {\cal{N}}^\prime_d$,
which can be written as
\begin{eqnarray}
 {\cal{N}}_d &=& \frac{v_2}{v_1} D_d - \frac{v_2}{\sqrt{2}}
\left( \frac{v_2}{v_1} +  \frac{v_1}{v_2}\right)  U^\dagger_{dL}
e^{i \alpha_2} \Gamma_2 \ U_{dR} -
\frac{v_2 \ v_3}{v_1\sqrt{2}} U^\dagger_{dL}e^{i \alpha_3} \Gamma_3  U_{dR}\\
  {\cal{N}}^\prime_d & = &  D_d -
\frac{v_3 - x}{\sqrt{2}} \  U^\dagger_{dL}
e^{i \alpha_3} \Gamma_3 \ U_{dR}
\end{eqnarray}
These expressions are general and can be particularised for our model.

\subsubsection{The  $(v, 0, 0)$ case}

In this case the matrices $\Gamma_1$, $\Gamma_2$ and $\Gamma_3$ are of the form
\begin{equation}
\Gamma_1 = 
\begin{pmatrix}
y_1^d & 0 & 0\\
0 & y_2^d & 0 \\
0 & 0 & y_3
\end{pmatrix}, \quad \Gamma_2 = \begin{pmatrix} 0_{3\times3} \end{pmatrix}, \quad \Gamma_3 = \begin{pmatrix} 0_{3\times3} \end{pmatrix},
\end{equation}
with similar expressions for the $\Delta$ matrices. The mass matrices of the up and down quarks sectors are diagonal, therefore $U_L$ and $U_R$ are identity matrices. In this case it is clear
from Eqs.~\eqref{Nd_general}--\eqref{Nup_general} that there are no FCNCs mediated by $R$, $I$, $R^\prime$ and $I^\prime$. In fact $\mathcal{N}_d$ and $\mathcal{N}_u$ are equal to zero while $\mathcal{N}^\prime_d$ and $\mathcal{N}^\prime_u$ are different from zero and diagonal.

\subsubsection{The $(v, v, v)$ case}

In this case the mass matrices in the down and up sectors have the structure given by Eq.~\eqref{eq:yuk111}
and the $\Gamma$ matrices are of the form
\begin{equation}
\Gamma_1 = 
\begin{pmatrix}
y_1^d & 0 & 0\\
y_2^d & 0 & 0 \\
y_3^d & 0 & 0
\end{pmatrix},
\quad
\Gamma_2 = 
\begin{pmatrix}
0 & y_1^d & 0\\
0 & \omega^2 y_2^d & 0 \\
0 & \omega y_3^d & 0
\end{pmatrix},
\quad
\Gamma_3 = 
\begin{pmatrix}
0 & 0 & y_1^d \\
0 & 0 & \omega y_2^d \\
0 & 0 & \omega^2 y_3
\end{pmatrix}
\end{equation}
and similarly for the matrices $\Delta$.
In this case the diagonalisation equation is
\begin{equation}
U^\dagger_L \ M \ U_R = \mathcal{D}
\end{equation}
and requires
\begin{equation}
U_L = \mathds{1}, \quad U_R = \frac{1}{\sqrt 3}
\begin{pmatrix}
1 & 1 & 1 \\
1 & \omega & \omega^2 \\
1 & \omega^2 & \omega
\end{pmatrix}.
\end{equation}
Eqs.~\eqref{Nd_general} and \eqref{Ndp_general} can be rewritten as
\begin{align}
\mathcal{N}_d &= \frac{1}{\sqrt{2}} U_{d_L}^\dagger v \left(\Gamma_1 + \Gamma_2 + \Gamma_3 - 2 \Gamma_2 - \Gamma_3 \right) U_{d_R} \\
&= \frac{1}{\sqrt{2}} D_d - \frac{v}{\sqrt{2}} \left( 2 \Gamma_2 + \Gamma_3 \right) U_{d_R},
\\
 {\cal{N}}^\prime_d  &= \frac{1}{\sqrt{2}} U_{d_L}^\dagger v \left(\Gamma_1 + \Gamma_2 + \Gamma_3 - 3 \Gamma_3 \right) U_{d_R} \\
&= \frac{1}{\sqrt{2}} D_d - \frac{v}{\sqrt{2}} \left( 3 \Gamma_3 \right) U_{d_R}
\end{align}
and similarly for the up sector. It can be readily verified that in both cases there is cancellation of the diagonal terms whilst the non-diagonal terms are in general different from zero. This means that there are FCNCs mediated by $R$, $I$, $R^\prime$ and $I^\prime$ but there are no flavour diagonal couplings to these scalars. As a result $b \ \rightarrow \  s \ \gamma$ can only be mediated by neutral Higgses $R$, $I$, $R^\prime$ and $I^\prime$ (which we choose to be heavier than the SM-like one) via diagrams with two flavour changing neutral vertices. The quark inside the loop will have to be the down quark and a suppression factor of its mass square divided by the square of the mass of the heavy scalar will occur.

Notice that $R$, $I$, $R^\prime$ and $I^\prime$ may not yet be the physical fields but $h^{\prime}$ already is and therefore does not mix with them.

\subsubsection{The $(\omega v, v, v)$ case}

In this case the matrices $\Gamma_1$, $\Gamma_2$ and $\Gamma_3$ can be read of Eq.~\eqref{eq:Yd}
 and are of the form
\begin{equation}
\Gamma_1 = 
\begin{pmatrix}
y_1^d & 0 & 0\\
0 & y_2^d & 0 \\
0 & 0 & y_3
\end{pmatrix}, \quad
\Gamma_2 = 
\begin{pmatrix} 
0 & y_1^d & 0 \\
0 & 0 & y_2^d  \\
y_3^d & 0 & 0  
\end{pmatrix}, \quad
\Gamma_3 = \begin{pmatrix}
0 & 0 & y_1^d \\
y_2^d & 0 & 0 \\
0 & y_3^d & 0
\end{pmatrix}.
\end{equation}
Now the diagonalisation equation requires
\begin{equation}
U_L = \mathds{1}, \quad U_R = \frac{1}{\sqrt 3}
\begin{pmatrix}
\omega^2 & 1 & 1 \\
1 & \omega^2 & 1 \\
1 & 1 & \omega^2
\end{pmatrix}.
\end{equation}
In this case we have for ${\cal{N}}_d$
\begin{eqnarray}
 {\cal{N}}_d &=& \frac{1}{\sqrt 2} \left( v \Gamma_1 - \omega v \Gamma_2 \right) U_{dR}\\
& = &
\frac{1}{\sqrt 2} \frac{1}{\sqrt 3} v
\begin{pmatrix} 
y_1^d \omega^2 - y_1^d \omega & 0 & y_1^d (1- \omega) \\
y_2^d (1- \omega) &  y_2^d \omega^2 - y_2^d \omega  & 0  \\
0 & y_3^d (1- \omega) & y_3^d \omega^2 - y_3^d \omega   
\end{pmatrix}. 
\end{eqnarray}
The (23) entry would correspond to a scalar mediated $b$ to $s$  transition, which is 
forbidden in this case. In what concerns $ {\cal{N}}^\prime_d$ we now have
\begin{eqnarray}
 {\cal{N}}^\prime_d &=& \frac{1}{\sqrt 2} v \left( \omega \Gamma_1 + \Gamma_2 - 2 \Gamma_3
\right) U_{dR}\\
& = &
\frac{1}{\sqrt 2} \frac{1}{\sqrt 3} v
\begin{pmatrix} 
0 & - 3 y_1^d  & -3 \omega^2 y_1^d  \\
-3 \omega^2 y_2^d  &  0 &  - 3 y_2^d \\
- 3 y_3^d & -3 \omega^2 y_3^d  & 0  
\end{pmatrix}.
\end{eqnarray}
In this matrix there are no flavour diagonal couplings and therefore $R^\prime$ and $I^\prime$ 
can only mediate $b \rightarrow s \gamma$ transitions with two flavour changing neutral vertices.

For the up quark mass matrix we have
\begin{equation}
\Delta_1 = 
\begin{pmatrix}
y_1^u & 0 & 0\\
0 & 0 & y_2^u \\
0 & y_3^u & 0
\end{pmatrix} \quad ; \quad 
\Delta_2 = 
\begin{pmatrix} 
0 & y_1^u & 0 \\
y_2^u & 0 & 0  \\
0 & 0 & y_3^u  
\end{pmatrix} \quad ; \quad 
\Delta_3 = \begin{pmatrix}
0 & 0 & y_1^u \\
0 & y_2^u & 0 \\
y_3^u & 0 & 0
\end{pmatrix}  \,.
\end{equation}
Now the diagonalisation equation requires
\begin{equation}
U_L = \mathds{1} \quad , \quad U_R = \frac{1}{\sqrt 3}
\begin{pmatrix}
\omega & 1 & 1 \\
1 & 1 & \omega \\
1 & \omega & 1
\end{pmatrix}
\end{equation}
and  ${\cal{N}}_u$ and $ {\cal{N}}^\prime_u$ are of the form
\begin{eqnarray}
 {\cal{N}}_u &=&
\frac{1}{\sqrt 2} \frac{1}{\sqrt 3} v
\begin{pmatrix} 
(\omega -\omega^2) y^u_1 & (1-\omega^2) y^u_1 & 0 \\
0 & (\omega -\omega^2) y^u_2  & (1-\omega^2) y^u_2  \\
(1-\omega^2) y^u_3  & 0 &(\omega -\omega^2) y^u_3  
\end{pmatrix},  \\
{\cal{N}}^\prime_u &=& \frac{1}{\sqrt 2} \frac{1}{\sqrt 3} v
\begin{pmatrix} 
0 & -3 y^u_1 \omega &  -3 y^u_1\\
-3 y^u_2 & 0  & -3 y^u_2 \omega  \\
 -3 y^u_3 \omega & -3 y^u_3 & 0  
\end{pmatrix} .
\end{eqnarray}
There are several similarities in the structure of  ${\cal{N}}_u$,  $ {\cal{N}}^\prime_u$ and of 
 ${\cal{N}}_d$ and $ {\cal{N}}^\prime_d$.

%% file: tex/phenomenology.tex
\section{Phenomenology of the model \label{sec:pheno}}

With the minima and respective Yukawa couplings described in the previous Section, we now proceed with a phenomenological analysis of each case.
For each vacuum choice we scan over 3 independent masses while fixing the SM-like Higgs boson mass $m_h$ to 125.25 GeV~\cite{ParticleDataGroup:2020ssz}.
For numerical analysis we use a \sa generated \texttt{SPheno} \cite{Porod:2003um,Porod:2011nf} spectrum generator to compute $STU$ parameters \cite{Peskin:1990zt,Marciano:1990dp,Kennedy:1990ib} and unitarity limit \cite{Goodsell:2018tti}.
In the case of $STU$ parameters we look for a region of masses within $3\sigma$ from best fit values of \cite{ParticleDataGroup:2020ssz}
\begin{align}
  S =& -0.01 \pm 0.10\\
  T =& \mathbin{\hphantom{-}} 0.03 \pm 0.12\\
  U =& \mathbin{\hphantom{-}} 0.02 \pm 0.11
\end{align}
While checking if unitarity limit is fulfilled we do take into account also finite scattering energy contributions, i.e. also contributions to scattering amplitudes from trilinear couplings (see Appendix~\ref{app:spheno_setup} for details regarding \texttt{SPheno} setting).
Finally, the Higgs sector is checked against experimental constraints using \texttt{HiggsBounds} \textit{v5.10.2} \cite{Bechtle:2008jh}, \texttt{HiggsSignals}~\textit{v2.6.2}~\cite{Bechtle:2013xfa} and also directly against experimental results if the above mentioned codes do not include a relevant analysis.

Because neither the module computing low energy observables nor the one computing unitarity constraints work with complex VEVs, in case of the $(\omega v, v, v)$ vacuum we use a rephasing freedom and analyse an equivalent case of a $(v, v, v)$ vacuum with the replacement $\Phi_1 \to \omega \Phi_1$.\footnote{To the \texttt{arXiv} version of this work we attach the rephased (and not the original) model.}

Finally, we note that the point of view of figures shown below differs from case to case and was chosen to better show non-trivial features of allowed parameter space regions.
In all cases charged Higgs bosons are mass degenerate (see Sec.~\ref{sec:model}) and their common mass $m_{H^\pm_1} = m_{H^\pm_2}$ is given on the vertical axis.
Horizontally, we show the 2 independent masses describing masses of the 4 remaining, pairwise degenerate, neutral states.

\subsection{The $(v,0,0)$ case}

\begin{figure}
  \subcaptionbox{$T$\label{fig:100_T}}[0.5\linewidth]{\raisebox{5mm}{\includegraphics[width=0.5\textwidth]{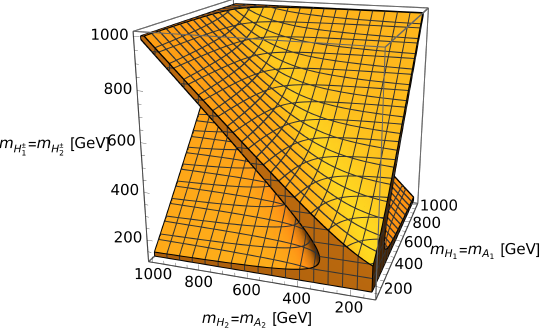}}}
  \subcaptionbox{unitarity\label{fig:100_unitarity}}[0.5\linewidth]{\includegraphics[width=0.41\textwidth]{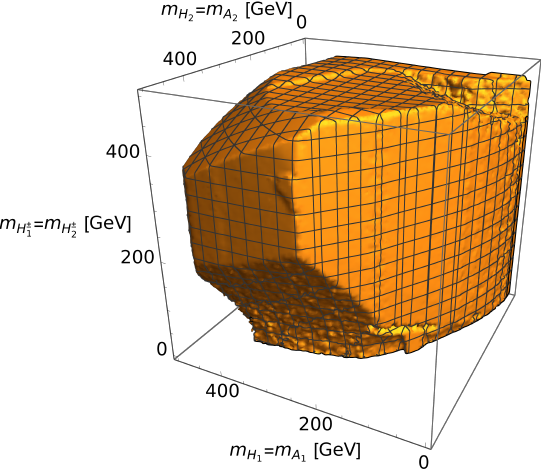}}
  \caption{
    Regions allowed by the $T$-parameter at 99\% level (a) and unitarity (b) for the $(v,0,0)$ vacuum in the $m_{H_1} = m_{A_1}$ \textit{vs.} $m_{H_2} = m_{A_2}$ \textit{vs.} $m_{H_1^\pm} = m_{H_2^\pm}$ space.
    The 4th free parameter, the mass of the SM-like Higgs $m_h$, is set to 125.25~GeV.
    On the left panel we additionally assume that all masses are larger than $m_h$ while on the right one that they are larger than 1 GeV.
  }
  \label{fig:100}
\end{figure}

In Fig.\ \ref{fig:100} we show masses allowed by the $T$-parameter and the unitarity constraint for the $(v,0,0)$ vacuum.
In this case pseudoscalars $A_1$ and $A_2$ are, pairwise, mass degenerate with $H_1$ and $H_2$, i.e. $m_{H_1} = m_{A_1}$, $m_{H_2} = m_{A_2}$.
Those masses are given in the horizontal planes of plots.
$S$ and $U$ parameters are not constraining at all within the shown range of masses.
This will also be the case for the remaining vacua.

As seen in Fig.~\ref{fig:100_T} the $T$-parameter forces either $m_{H_1} = m_{A_1} \approx m_{H^\pm}$ or $m_{H_2} = m_{A_2} \approx m_{H^\pm}$ or both.
This is similar as in the case of the 2HDM as shown for example in \cite{Haller:2018nnx}.
Meanwhile, the unitarity constraints lead to the conclusion that the masses of $H^\pm$ and neutral Higgses must be limited to $\lesssim 500$ GeV.

As far as direct experimental limits are concerned, non-SM Higgses decay to pairs consisting of an another Higgs and a gauge boson, or to fermions in the flavour-violating manner (see Appendix~\ref{sec:br} for an example decay pattern).
Limits on neutral beyond SM Higgses are avoided because of both their non-standard decay patterns and because the leading production channels of $H$'s and $A$'s here are $\bar{b} q$ + h.c., where $q = d, s$ and not the $gg$ fusion.
Vector boson fusion (VBF) and Higgsstrahlung is also forbidden as beyond the SM Higgses do not have VEVs and therefore lack triple $VVH$ couplings.
Therefore typical limits, like the observed limit on the BR$(H \to e \mu) < 6.1\cdot10^{-5}$ from \cite{ATLAS:2019old}, do not apply as they assume SM-like Higgs boson with mass of 125 GeV and SM production channels.
This allows them to evade experimental limits.
No points are excluded by \hb, while \hs only requires that $\text{min}(m_{H_1}, \, m_{H_2}, \,  m_{H^\pm}) > m_h/2$ as otherwise 2 body decays of a SM-like Higgs to neutral or charged (or both) Higgses are open.\footnote{We point out that due to a problem with tagging VEV-less fields as `Higgses` in \sa, in the case of $(v,0,0)$ vacuum fields $H_i$ are not checked against experimental limits in \hb and \hs. There are checked in the remaining cases though.
Since in those cases the only constraints on $H_i$ come from them being a decay products of $h$ (which is already taken into account in the $(v,0,0)$ case) and $H_i$ decay patters do not differ between cases, we conclude that checking $H_i$ against \hs and \hb in the $(v,0,0)$ case would not lead to any new constraints.}
The example point fulfilling all constrains is $m_{H_1} = m_{A_1} = 134.2$ GeV, $m_{H_2} = m_{A_2} = 139.3$ GeV and $m_{H^\pm} = 165.6$ GeV, giving $T = 0.03$ and passing the unitarity check.

With the \sa interface to \hb the LEP production cross sections of charged Higgses are not considered, only their partial widths (see comments in \cite{sarah:HB}).
\hs puts a limit of $m_H^\pm \gtrsim 95$ GeV.
We emphasize though that standard limits do not apply in this case either.
The LEP limit constrains the $H^+ H^-$ production if $\text{BR}(H^+ \to c \bar{s}) + \text{BR}(H^+ \to \tau^+ \nu)  \approx 1$.
Since there are no diagonal couplings of $H^\pm$ to quarks, those limits do not apply in our case as $H^+_i$ decay almost exclusively to $\bar{b}u$ and $\bar{b}c$.
Similarly, the limits from LHC, which come from $t \to b H^+$ decay or an associated $t\bar{b} H^-$ + h.c.\ production   (as in \cite{ATLAS:2018gfm}), do not  apply either.

\subsection{The $(v,v,v)$ case}

\begin{figure}
  \subcaptionbox{$T$\label{fig:111_T}}[0.5\linewidth]{\raisebox{0mm}{\includegraphics[width=0.5\textwidth]{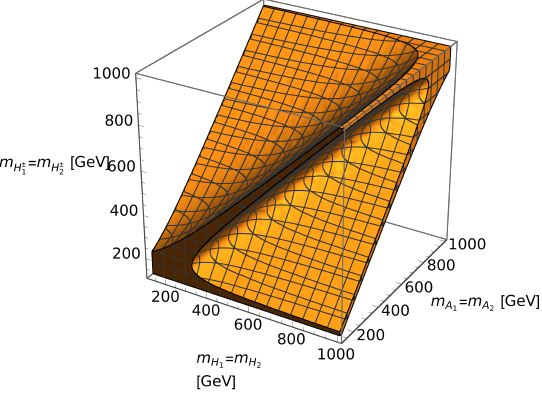}}}
  \subcaptionbox{unitarity\label{fig:111_unitarity}}[0.5\linewidth]{\includegraphics[width=0.41\textwidth]{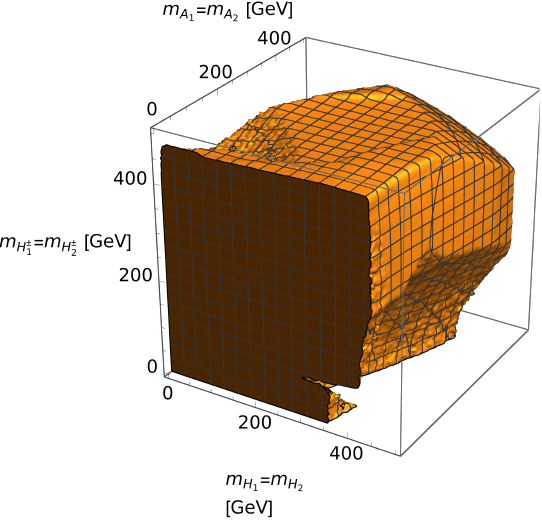}}
  \caption{
      Regions allowed by the $T$-parameter at 99\% level (a) and unitarity (b) for the $(v,v,v)$ vacuum in the $m_{H_1} = m_{H_2}$ \textit{vs.} $m_{A_1} = m_{A_2}$ \textit{vs.} $m_{H_1^\pm} = m_{H_2^\pm}$ space.
    The 4th free parameter, the mass of the SM-like Higgs $m_h$, is set to 125.25~GeV.
    On the left panel we additionally assume that all masses are larger than $m_h$ while on the right one that they are larger than 1 GeV.
  }
  \label{fig:111}
\end{figure}

In this case  the scalar and pseudoscalar Higgses have, separately, common masses:  $m_{H_1}=m_{H_2}$ and $m_{A_1}=m_{A_2}$.
The $T$-parameter allowed region in Fig.~\ref{fig:111_T} has the same features as the one in Fig.~\ref{fig:100_T}, giving analogous relations between masses of $SU(2)_L$ doublet components.
Unitarity also puts a similar limit on the masses requiring them to be smaller than about 500 GeV (Fig.~\ref{fig:111_unitarity}).
No points are excluded by \hb and, similarly as before, \hs limits masses to $\text{min}(m_{H_1}, m_{A_1}) > m_h/2$ and
$m_{H^\pm} \gtrsim 95$~GeV.
The same arguments hold as in the case of $(v,0,0)$, including lack of VBF and Higgsstrahlung processes.
The lack of VBF and Higgsstrahlung processes for $(v,v,v)$ case follows from the fact, as explained in Sec.~\ref{subsec:FCNC}, that the SM-like Higgs does not change between the Higgs basis and mass eigenstates meaning that non-SM Higgses do not possess a VEV -- the same as in the $(v,0,0)$ case.

The example point fulfilling unitarity and $T$-parameter constraint ($T=0.03$) is $m_H = 198.6$~GeV, $m_A = 202.4$~GeV, $m_{H^\pm} = 172.3$~GeV.

\subsection{The $(\omega v,v,v)$ case}

\begin{figure}
  \subcaptionbox{$T$\label{fig:omega1_T}}[0.5\linewidth]{\raisebox{2mm}{\includegraphics[width=0.42\textwidth]{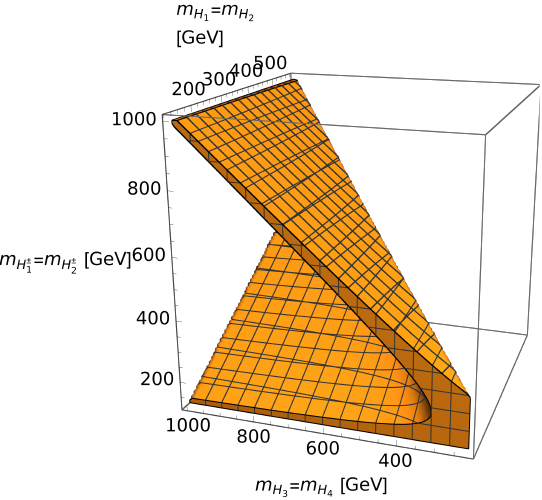}}}
  \subcaptionbox{unitarity\label{fig:omega1_unitarity}}[0.5\linewidth]{\includegraphics[width=0.5\textwidth]{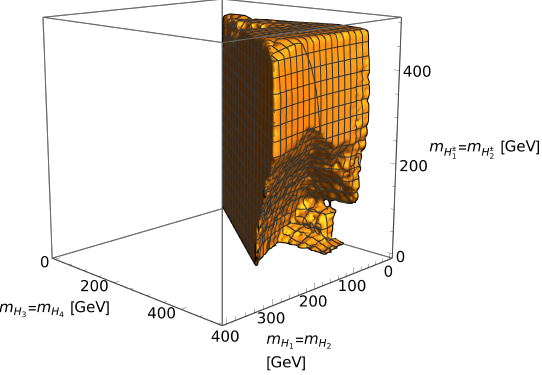}}
  \caption{
    Regions allowed by the $T$-parameter at 99\% level (a) and unitarity (b) for the $(\omega v,v,v)$ vacuum in the $m_{H_1} = m_{H_2}$ \textit{vs.} $m_{H_3} = m_{H_4}$ \textit{vs.} $m_{H_1^\pm} = m_{H_2^\pm}$ space.
    The 4th free parameter, the mass of the SM-like Higgs $m_h$, is set to 125.25~GeV.
    On the left panel we additionally assume that all masses are larger than $m_h$ while on the right one that they are larger than 1 GeV.
  }
  \label{fig:omega}
\end{figure}

In this case CP is ill-defined so we label all neutral mass eigenstates beyond the SM-like state $h$ as $H_i$ ($i=1\ldots4$).
We find pairwise mass degeneracies, with $H_1$, $H_2$ sharing one mass and $H_3$ and $H_4$ sharing another one.
As explained in Sec.~\ref{sec:omega} there is a double degeneracy in solution for potential parameters for a given set of masses.
Qualitatively, the allowed region of masses is the same in both cases therefore in Fig.~\ref{fig:omega} we only plot one of the solutions.
Most of the features of Fig.~\ref{fig:omega} where already present in Figs.~\ref{fig:100} and \ref{fig:111} with the distinctive feature in the current case being mainly the condition $m_{H_3} > \sqrt{3} m_{H_1}$.

An example point passing all phenomenological constraints is: $m_{H_1} = m_{H_2} = 158.6$ GeV, $m_{H_3} = m_{H_4} = 314.3$ GeV, $m_{H^\pm} = 149.7$ GeV with $T = 0.03$, fulfilled unitarity bound and not excluded by \hb and \hs.

\subsection{Comment on unitarity}

As seen from the discussion in this section, the unitarity provides a very strong constraint, limiting all masses to $\lesssim 500$ GeV irrespectively of the chosen vacuum.
This is relatively easy to understand.
As masses increase, so must the absolute values of couplings as the mass scale is fixed by the EWSB scale.
Qualitatively, regions allowed by unitarity are roughly given by $max \{|s|, |r_1|, |r_2|, |d|\} \lesssim \pi$.
We emphasise though that in our analysis we consider also finite energy contributions to the scattering matrix, making it sensitive also to 3-point interactions originating from Eq.~\eqref{V27H} after the electroweak symmetry breaking.

%% file: tex/conclusions.tex
\section{Conclusions \label{sec:Conc}}

We have described an extension of the Standard Model with 3 Higgs doublets and  the potential invariant under the $\Delta(27)$ symmetry.
After covering the known tree-level minima of this potential, we gave formulas for the physical parameters (masses) in terms of the potential parameters and then proceed to constrain the parameter space of the model through $STU$, unitarity as well as certain flavour violation processes. We concluded that unitarity
provides a rather strong constraint and keeps the scalar masses below around 500 GeV.
We considered assignments of the fermions under the $\Delta(27)$ symmetry that give Yukawa structures leading to distinct masses for each generation and the unit CKM matrix (which we consider a reasonable leading order approximation).
Based on that toy model of Yukawa couplings, we were able to test contributions of the additional scalars to flavour violating processes. Using $b \rightarrow s \gamma$ as an example, we concluded that flavour violating processes are tamed by the imposed discrete symmetry. The Higgs mediated neutral flavour changing  contributions are suppressed by ratios of the mass of the down-type quark in the loop to the mass of the heavy Higgs. In our model, whenever such contributions are present, two flavour changing neutral vertices are needed implying that the quark in the loop is the down quark, which is extremely light compared to the mass of the heavy Higgs.

%% file: tex/acknowledgments.tex
\section*{Acknowledgments}

Work supported by the Polish National Science Centre HARMONIA grant under contract UMO-2015/18/M/ST2/00518 (2016-2021) and by the Funda{\c{c}}{\~{a}}o para a Ci{\^{e}}ncia
e a Tecnologia (FCT, Portugal) through the projects CFTP-FCT Unit 777
(UIDB/00777/2020 and UIDP/00777/2020), PTDC/FIS-PAR/29436/2017, CERN/FIS-PAR/0004/2019 and CERN/FIS-PAR/0008/2019, which are partially funded through POCTI (FEDER), COMPETE, QREN and EU.
IdMV acknowledges funding from the Funda\c{c}\~{a}o para a Ci\^{e}ncia e a Tecnologia (FCT, Portugal) through the
contract IF/00816/2015.
WK was supported in part by the German Research Foundation (DFG) under grant number STO 876/2-2 and by the National Science Centre (Poland) under the research grant 2020/38/E/ST2/00126.

IdMV and MNR thank the University of Warsaw for hospitality during their visits supported by the HARMONIA grant.
JK thanks the Theory Division of CERN and JK and WK thank CFTP, Lisbon for hospitality during the final stage of this work.

The authors are grateful to the Centre for Information Services and High Performance Computing [Zentrum für Informationsdienste und Hochleistungsrechnen (ZIH)] TU Dresden for providing its facilities for high throughput calculations.

%% file: tex/D27group.tex
\section{Group theory of $\Delta(27)$ \label{app:D27}}

$\Delta(27)$ is a discrete subgroup of $SU(3)$ with two complex 3-dimensional and 9 distinct 1-dimensional irreducible representations. Here we label one of the 3-dimensional ones as the anti-triplet $\bar{\mathbf{3}}$, the other as the triplet $\mathbf{3}$, and label the singlets as $\mathbf{1}_{k,l}$ ($k,l=0,1,2$) --- where the labels $k,l$ denote the transformation properties under the order 3 generators $c$, $d$ of the group ($c^3=1$ and $d^3=1$).
The generators for 1-dimensional representations are simply the phases (powers of $\omega\equiv e^{i\frac{2\pi }{3}}$). In a convenient basis, which we use throughout, for the triplet and anti-triplet representation we have
\begin{align}
c_{3 , \bar{3}} =
\begin{pmatrix}
0 & 1 & 0 \\ 
0 & 0 & 1 \\
1 & 0 & 0
\end{pmatrix}
\end{align}

\begin{align}
d_3 =
\begin{pmatrix}
1 & 0 & 0 \\ 
0 & \omega & 0 \\
0 & 0 & \omega^2
\end{pmatrix}
\end{align}

\begin{align}
d_{\bar{3}} =
\begin{pmatrix}
1 & 0 & 0 \\ 
0 & \omega^2 & 0 \\
0 & 0 & \omega
\end{pmatrix}
\end{align}

With this notation, as the generators of the group are of order 3, the trivial singlet is obtained from products of singlets where the labels add up to zero modulo 3. The product of a triplet and anti-triplet results in all nine singlets whereas the product of two anti-triplets yields three triplets and vice-versa as follows
\begin{eqnarray}
\mathbf{3}\otimes \mathbf{3} &=&\bar{\mathbf{3}}_{S_{1}}\oplus 
\bar{\mathbf{3}}_{S_{2}}\oplus \bar{\mathbf{3}}_{A}  \notag \\
\bar{\mathbf{3}}\otimes \bar{\mathbf{3}} &=&\mathbf{3}%
_{S_{1}}\oplus \mathbf{3}_{S_{2}}\oplus \mathbf{3}_{A}  \notag \\
\mathbf{3}\otimes \bar{\mathbf{3}} &=&\sum_{r=0}^{2}\mathbf{1}%
_{r,0}\oplus \sum_{r=0}^{2}\mathbf{1}_{r,1}\oplus \sum_{r=0}^{2}\mathbf{1}%
_{r,2}  \notag \\
\mathbf{1}_{k,\ell }\otimes \mathbf{1}_{k^{\prime },\ell ^{\prime }} &=&%
\mathbf{1}_{k+k^{\prime}~mod~3,\ell +\ell^{\prime}~mod~3}
\end{eqnarray}%
The specific composition rules for the (anti-)triplet products are
\begin{eqnarray}
\left( \mathbf{3}\otimes \mathbf{3}\right) _{\bar{\mathbf{3}}_{S_{1}}}
&=&\left( x_{1}y_{1},x_{2}y_{2},x_{3}y_{3}\right) ,  \notag
\label{triplet-vectors} \\
\left( \mathbf{3}\otimes \mathbf{3}\right) _{\bar{\mathbf{3}}_{S_{2}}}
&=&\frac{1}{2}\left(
x_{2}y_{3}+x_{3}y_{2},x_{3}y_{1}+x_{1}y_{3},x_{1}y_{2}+x_{2}y_{1}\right) , 
\notag \\
\left( \mathbf{3}\otimes \mathbf{3}\right) _{\bar{\mathbf{3}}_{A}} &=&%
\frac{1}{2}\left(
x_{2}y_{3}-x_{3}y_{2},x_{3}y_{1}-x_{1}y_{3},x_{1}y_{2}-x_{2}y_{1}\right) , 
\notag \\
\left( \mathbf{3}\otimes \bar{\mathbf{3}}\right) _{\mathbf{1}_{r,0}}
&=&x_{1}y_{1}+\omega ^{2r}x_{2}y_{2}+\omega ^{r}x_{3}y_{3},  \notag \\
\left( \mathbf{3}\otimes \bar{\mathbf{3}}\right) _{\mathbf{1}_{r,1}}
&=&x_{1}y_{2}+\omega ^{2r}x_{2}y_{3}+\omega ^{r}x_{3}y_{1},  \notag \\
\left( \mathbf{3}\otimes \bar{\mathbf{3}}\right) _{\mathbf{1}_{r,2}}
&=&x_{1}y_{3}+\omega ^{2r}x_{2}y_{1}+\omega ^{r}x_{3}y_{2},
\end{eqnarray}%
where $r=0,1,2$, and $\left( x_{1},x_{2},x_{3}\right) $ and $\left(
y_{1},y_{2},y_{3}\right) $ are the components of the (anti-)triplets in the product. Care should be taken in the product of a triplet and anti-triplet as the ordering is relevant.
The subscript for the product of a triplet and an anti-triplet resulting in a singlet specifies which of the singlets it transforms as, whereas the product of two triplets (or two anti-triplets) results in distinct anti-triplets (or triplets), where the subscripts denote if it is a symmetric product ($S_1$, $S_2$) or an anti-symmetric one ($A$).

%% file: tex/spheno_setup.tex
\section{\texttt{SPheno} setup}
\label{app:spheno_setup}
For numerical analysis we use \texttt{SPheno} in the \texttt{OnlyLowEnergySPheno} setup.
Higgs boson masses are computed purely at the tree level and parameter points are specified in terms of physical Higgs masses, with scalar potential parameters extracted using Eqs.~\eqref{eq:pot_pars_100_1}--\eqref{eq:pot_pars_100_2}, \eqref{eq:pot_pars_111_1}--\eqref{eq:pot_pars_111_2} or \eqref{eq:pot_pars_o11_1}--\eqref{eq:pot_pars_o11_2}.
We equate 3HDM VEVs (with obvious numerical prefactors) to the SM VEV fixed to $v_{\text{SM}} \equiv \left(\sqrt{2} G_F \right)^{-1/2}$, with $G_F = 1.1663787 \cdot 10^{-5}$ GeV$^{-2}$.
No RGE running is performed, so through calculation of all observables (unitarity, $STU$, etc.) the same values of $s$, $r_1$, $r_2$ and $d$ are used.
There is however a small difference in gauge and Yukawa couplings used in different parts of the code due to how \texttt{SPheno} extracts them from SM input parameters.

We use the following settings in \texttt{SPheno} to check the unitarity constraints

\begin{lstlisting}[language=SLHA]
BLOCK SPhenoInput
  440 1               # Tree-level unitarity constraints (limit s->infinity)
  441 1               # Full tree-level unitarity constraints
  442 1.              # sqrt(s_min)
  443 5000.           # sqrt(s_max)
  444 -1000           # steps
  445 0               # running
  445 2               # Cut-Level for T/U poles
\end{lstlisting}

Technically, as for the (finite) maximal energy of 5 TeV, occasionally the maximal scattering eigenvalue $\lambda_{\text{max}}^{\text{finite}~S} < \lambda_{\text{max}}^{S \to \inf}$ we plot the region where $max(\lambda_{\text{max}}^{\text{finite}~S}, \lambda_{\text{max}}^{S \to \inf}) < \tfrac{1}{2}$. Here $\lambda_{\text{max}}^{\text{finite}~S}$ $(\lambda_{\text{max}}^{S \to \inf})$ is the eigenvalue as given in block \texttt{TREELEVELUNITARITYwTRILINEARS} (\texttt{TREELEVELUNITARITY}) of \texttt{SPheno}'s \texttt{SLHA}~\cite{Skands:2003cj,Allanach:2008qq,Goodsell:2018tti} output.

%% file: tex/decays.tex
\section{Example decay patterns of non-SM Higgs bosons}
\label{sec:br}
Example of decay patters of non-SM Higgses for the $(v,0,0)$ vacuum and $m_{H_1} = 200$ GeV, $m_{H_2} = 300$ GeV and $m_{H_2^\pm} = 400$ GeV as given by the created 3HDM \texttt{SPheno} spectrum generator.
In the extract of the \texttt{SLHA} output below, \texttt{H0\_2}, \texttt{A0\_2}, \texttt{Hp\_2}  correspond to $H_2$, $A_1$ and $H^\pm_1$, respectively.
For other particles standard \texttt{SPheno} symbols and indexing are used.

\begin{lstlisting}
DECAY        45     7.97690384E+00   # H0_2
#    BR                NDA      ID1      ID2
     1.29114182E-02    2           36         23   # BR(H0_2 -> A0_2 VZ )
     3.23331057E-04    2           -1          5   # BR(H0_2 -> Fd_1^* Fd_3 )
     3.23497669E-04    2           -3          5   # BR(H0_2 -> Fd_2^* Fd_3 )
     3.23331057E-04    2           -5          1   # BR(H0_2 -> Fd_3^* Fd_1 )
     3.23497669E-04    2           -5          3   # BR(H0_2 -> Fd_3^* Fd_2 )
     2.46405648E-01    2           -2          6   # BR(H0_2 -> Fu_1^* Fu_3 )
     2.46424262E-01    2           -4          6   # BR(H0_2 -> Fu_2^* Fu_3 )
     2.46405648E-01    2           -6          2   # BR(H0_2 -> Fu_3^* Fu_1 )
     2.46424262E-01    2           -6          4   # BR(H0_2 -> Fu_3^* Fu_2 )
DECAY        36     7.60203352E-01   # A0_2
#    BR                NDA      ID1      ID2
     9.65338080E-04    2           -1          5   # BR(A0_2 -> Fd_1^* Fd_3 )
     9.65695420E-04    2           -3          5   # BR(A0_2 -> Fd_2^* Fd_3 )
     9.65338080E-04    2           -5          1   # BR(A0_2 -> Fd_3^* Fd_1 )
     9.65695420E-04    2           -5          3   # BR(A0_2 -> Fd_3^* Fd_2 )
     1.36165292E-04    2          -11         15   # BR(A0_2 -> Fe_1^* Fe_3 )
     1.36636215E-04    2          -13         15   # BR(A0_2 -> Fe_2^* Fe_3 )
     1.36165292E-04    2          -15         11   # BR(A0_2 -> Fe_3^* Fe_1 )
     1.36636215E-04    2          -15         13   # BR(A0_2 -> Fe_3^* Fe_2 )
     2.49066556E-01    2           -2          6   # BR(A0_2 -> Fu_1^* Fu_3 )
     2.48685848E-01    2           -4          6   # BR(A0_2 -> Fu_2^* Fu_3 )
     2.49066556E-01    2           -6          2   # BR(A0_2 -> Fu_3^* Fu_1 )
     2.48685848E-01    2           -6          4   # BR(A0_2 -> Fu_3^* Fu_2 )
DECAY        37     3.06760690E+01   # Hp_2
#    BR                NDA      ID1      ID2
     1.11313431E-01    2           36         24   # BR(Hp_2 -> A0_2 VWp )
     6.11469833E-03    2           46         24   # BR(Hp_2 -> A0_3 VWp )
     2.96841032E-04    2           -1          6   # BR(Hp_2 -> Fd_1^* Fu_3 )
     7.64780553E-01    2           -5          4   # BR(Hp_2 -> Fd_3^* Fu_2 )
     1.11313431E-01    2           35         24   # BR(Hp_2 -> H0_1 VWp )
     6.11469833E-03    2           45         24   # BR(Hp_2 -> H0_2 VWp )
\end{lstlisting}
The entirety of the above output is attached to the \texttt{arXiv} version of this work.

%% file: 3HDMDelta27.bbl
\providecommand{\href}[2]{#2}\begingroup\raggedright\begin{thebibliography}{10}

\bibitem{Gunion:1989we}
J.F.~Gunion, H.E.~Haber, G.L.~Kane and S.~Dawson, \emph{The higgs hunter’s
  guide, vol. 80}, {\emph{Front. Phys} (2000) 1}.

\bibitem{Branco:2011iw}
G.C.~Branco, P.M.~Ferreira, L.~Lavoura, M.N.~Rebelo, M.~Sher and J.P.~Silva,
  \emph{{Theory and phenomenology of two-Higgs-doublet models}},
  \href{https://doi.org/10.1016/j.physrep.2012.02.002}{\emph{Phys. Rept.}
  {\bfseries 516} (2012) 1} [\href{https://arxiv.org/abs/1106.0034}{{\ttfamily
  1106.0034}}].

\bibitem{Ivanov:2012fp}
I.P.~Ivanov and E.~Vdovin, \emph{{Classification of finite reparametrization
  symmetry groups in the three-Higgs-doublet model}},
  \href{https://doi.org/10.1140/epjc/s10052-013-2309-x}{\emph{Eur. Phys. J. C}
  {\bfseries 73} (2013) 2309}
  [\href{https://arxiv.org/abs/1210.6553}{{\ttfamily 1210.6553}}].

\bibitem{Darvishi:2019dbh}
N.~Darvishi and A.~Pilaftsis, \emph{{Classifying Accidental Symmetries in
  Multi-Higgs Doublet Models}},
  \href{https://doi.org/10.1103/PhysRevD.101.095008}{\emph{Phys. Rev. D}
  {\bfseries 101} (2020) 095008}
  [\href{https://arxiv.org/abs/1912.00887}{{\ttfamily 1912.00887}}].

\bibitem{Darvishi:2021txa}
N.~Darvishi, M.R.~Masouminia and A.~Pilaftsis, \emph{{Maximally symmetric
  three-Higgs-doublet model}},
  \href{https://doi.org/10.1103/PhysRevD.104.115017}{\emph{Phys. Rev. D}
  {\bfseries 104} (2021) 115017}
  [\href{https://arxiv.org/abs/2106.03159}{{\ttfamily 2106.03159}}].

\bibitem{Toorop:2010ex}
R.~de~Adelhart~Toorop, F.~Bazzocchi, L.~Merlo and A.~Paris, \emph{{Constraining
  Flavour Symmetries At The EW Scale I: The A4 Higgs Potential}},
  \href{https://doi.org/10.1007/JHEP03(2011)035}{\emph{JHEP} {\bfseries 03}
  (2011) 035} [\href{https://arxiv.org/abs/1012.1791}{{\ttfamily 1012.1791}}].

\bibitem{Toorop:2010kt}
R.~de~Adelhart~Toorop, F.~Bazzocchi, L.~Merlo and A.~Paris, \emph{{Constraining
  Flavour Symmetries At The EW Scale II: The Fermion Processes}},
  \href{https://doi.org/10.1007/JHEP03(2011)040}{\emph{JHEP} {\bfseries 03}
  (2011) 040} [\href{https://arxiv.org/abs/1012.2091}{{\ttfamily 1012.2091}}].

\bibitem{Branco:1983tn}
G.C.~Branco, J.M.~Gerard and W.~Grimus, \emph{{Geometrical T-violation}},
  \href{https://doi.org/10.1016/0370-2693(84)92024-0}{\emph{Phys. Lett. B}
  {\bfseries 136} (1984) 383}.

\bibitem{deMedeirosVarzielas:2006fc}
I.~de~Medeiros~Varzielas, S.F.~King and G.G.~Ross, \emph{{Neutrino
  tri-bi-maximal mixing from a non-Abelian discrete family symmetry}},
  \href{https://doi.org/10.1016/j.physletb.2007.03.009}{\emph{Phys. Lett. B}
  {\bfseries 648} (2007) 201}
  [\href{https://arxiv.org/abs/hep-ph/0607045}{{\ttfamily hep-ph/0607045}}].

\bibitem{Ma:2006ip}
E.~Ma, \emph{{Neutrino Mass Matrix from Delta(27) Symmetry}},
  \href{https://doi.org/10.1142/S0217732306021190}{\emph{Mod. Phys. Lett. A}
  {\bfseries 21} (2006) 1917}
  [\href{https://arxiv.org/abs/hep-ph/0607056}{{\ttfamily hep-ph/0607056}}].

\bibitem{deMedeirosVarzielas:2011zw}
I.~de~Medeiros~Varzielas and D.~Emmanuel-Costa, \emph{{Geometrical CP
  Violation}}, \href{https://doi.org/10.1103/PhysRevD.84.117901}{\emph{Phys.
  Rev. D} {\bfseries 84} (2011) 117901}
  [\href{https://arxiv.org/abs/1106.5477}{{\ttfamily 1106.5477}}].

\bibitem{Varzielas:2012nn}
I.~de~Medeiros~Varzielas, D.~Emmanuel-Costa and P.~Leser, \emph{{Geometrical CP
  Violation from Non-Renormalisable Scalar Potentials}},
  \href{https://doi.org/10.1016/j.physletb.2012.08.008}{\emph{Phys. Lett. B}
  {\bfseries 716} (2012) 193}
  [\href{https://arxiv.org/abs/1204.3633}{{\ttfamily 1204.3633}}].

\bibitem{Bhattacharyya:2012pi}
G.~Bhattacharyya, I.~de~Medeiros~Varzielas and P.~Leser, \emph{{A common origin
  of fermion mixing and geometrical CP violation, and its test through Higgs
  physics at the LHC}},
  \href{https://doi.org/10.1103/PhysRevLett.109.241603}{\emph{Phys. Rev. Lett.}
  {\bfseries 109} (2012) 241603}
  [\href{https://arxiv.org/abs/1210.0545}{{\ttfamily 1210.0545}}].

\bibitem{Ferreira:2012ri}
P.M.~Ferreira, W.~Grimus, L.~Lavoura and P.O.~Ludl, \emph{{Maximal CP Violation
  in Lepton Mixing from a Model with Delta(27) flavour Symmetry}},
  \href{https://doi.org/10.1007/JHEP09(2012)128}{\emph{JHEP} {\bfseries 09}
  (2012) 128} [\href{https://arxiv.org/abs/1206.7072}{{\ttfamily 1206.7072}}].

\bibitem{Ma:2013xqa}
E.~Ma, \emph{{Neutrino Mixing and Geometric CP Violation with Delta(27)
  Symmetry}}, \href{https://doi.org/10.1016/j.physletb.2013.05.011}{\emph{Phys.
  Lett. B} {\bfseries 723} (2013) 161}
  [\href{https://arxiv.org/abs/1304.1603}{{\ttfamily 1304.1603}}].

\bibitem{Nishi:2013jqa}
C.C.~Nishi, \emph{{Generalized $CP$ symmetries in $\Delta(27)$ flavor models}},
  \href{https://doi.org/10.1103/PhysRevD.88.033010}{\emph{Phys. Rev. D}
  {\bfseries 88} (2013) 033010}
  [\href{https://arxiv.org/abs/1306.0877}{{\ttfamily 1306.0877}}].

\bibitem{Varzielas:2013sla}
I.~de~Medeiros~Varzielas and D.~Pidt, \emph{{Towards realistic models of quark
  masses with geometrical CP violation}},
  \href{https://doi.org/10.1088/0954-3899/41/2/025004}{\emph{J. Phys. G}
  {\bfseries 41} (2014) 025004}
  [\href{https://arxiv.org/abs/1307.0711}{{\ttfamily 1307.0711}}].

\bibitem{Aranda:2013gga}
A.~Aranda, C.~Bonilla, S.~Morisi, E.~Peinado and J.W.F.~Valle, \emph{{Dirac
  neutrinos from flavor symmetry}},
  \href{https://doi.org/10.1103/PhysRevD.89.033001}{\emph{Phys. Rev. D}
  {\bfseries 89} (2014) 033001}
  [\href{https://arxiv.org/abs/1307.3553}{{\ttfamily 1307.3553}}].

\bibitem{Varzielas:2013eta}
I.~de~Medeiros~Varzielas and D.~Pidt, \emph{{Geometrical CP violation with a
  complete fermion sector}},
  \href{https://doi.org/10.1007/JHEP11(2013)206}{\emph{JHEP} {\bfseries 11}
  (2013) 206} [\href{https://arxiv.org/abs/1307.6545}{{\ttfamily 1307.6545}}].

\bibitem{Harrison:2014jqa}
P.F.~Harrison, R.~Krishnan and W.G.~Scott, \emph{{Deviations from tribimaximal
  neutrino mixing using a model with $\Delta(27)$ symmetry}},
  \href{https://doi.org/10.1142/S0217751X1450095X}{\emph{Int. J. Mod. Phys. A}
  {\bfseries 29} (2014) 1450095}
  [\href{https://arxiv.org/abs/1406.2025}{{\ttfamily 1406.2025}}].

\bibitem{Ma:2014eka}
E.~Ma and A.~Natale, \emph{{Scotogenic $Z_2$ or $U(1)_D$ Model of Neutrino Mass
  with $\Delta(27)$ Symmetry}},
  \href{https://doi.org/10.1016/j.physletb.2014.05.070}{\emph{Phys. Lett. B}
  {\bfseries 734} (2014) 403}
  [\href{https://arxiv.org/abs/1403.6772}{{\ttfamily 1403.6772}}].

\bibitem{Fallbacher:2015rea}
M.~Fallbacher and A.~Trautner, \emph{{Symmetries of symmetries and geometrical
  CP violation}},
  \href{https://doi.org/10.1016/j.nuclphysb.2015.03.003}{\emph{Nucl. Phys. B}
  {\bfseries 894} (2015) 136}
  [\href{https://arxiv.org/abs/1502.01829}{{\ttfamily 1502.01829}}].

\bibitem{Abbas:2015zna}
M.~Abbas, S.~Khalil, A.~Rashed and A.~Sil, \emph{{Neutrino masses and deviation
  from tribimaximal mixing in \ensuremath{\Delta}(27) model with inverse seesaw
  mechanism}}, \href{https://doi.org/10.1103/PhysRevD.93.013018}{\emph{Phys.
  Rev. D} {\bfseries 93} (2016) 013018}
  [\href{https://arxiv.org/abs/1508.03727}{{\ttfamily 1508.03727}}].

\bibitem{Varzielas:2015aua}
I.~de~Medeiros~Varzielas, \emph{{$\Delta(27)$ family symmetry and neutrino
  mixing}}, \href{https://doi.org/10.1007/JHEP08(2015)157}{\emph{JHEP}
  {\bfseries 08} (2015) 157}
  [\href{https://arxiv.org/abs/1507.00338}{{\ttfamily 1507.00338}}].

\bibitem{Bjorkeroth:2015uou}
F.~Bj\"orkeroth, F.J.~de~Anda, I.~de~Medeiros~Varzielas and S.F.~King,
  \emph{{Towards a complete $\Delta(27) \times SO(10)$ SUSY GUT}},
  \href{https://doi.org/10.1103/PhysRevD.94.016006}{\emph{Phys. Rev. D}
  {\bfseries 94} (2016) 016006}
  [\href{https://arxiv.org/abs/1512.00850}{{\ttfamily 1512.00850}}].

\bibitem{Chen:2015jta}
P.~Chen, G.-J.~Ding, A.D.~Rojas, C.A.~Vaquera-Araujo and J.W.F.~Valle,
  \emph{{Warped flavor symmetry predictions for neutrino physics}},
  \href{https://doi.org/10.1007/JHEP01(2016)007}{\emph{JHEP} {\bfseries 01}
  (2016) 007} [\href{https://arxiv.org/abs/1509.06683}{{\ttfamily
  1509.06683}}].

\bibitem{Hernandez:2016eod}
A.E.~C\'arcamo~Hern\'andez, H.N.~Long and V.V.~Vien, \emph{{A 3-3-1 model with
  right-handed neutrinos based on the $\varDelta \left( 27\right) $ family
  symmetry}}, \href{https://doi.org/10.1140/epjc/s10052-016-4074-0}{\emph{Eur.
  Phys. J. C} {\bfseries 76} (2016) 242}
  [\href{https://arxiv.org/abs/1601.05062}{{\ttfamily 1601.05062}}].

\bibitem{CarcamoHernandez:2017owh}
A.E.~C\'arcamo~Hern\'andez, S.~Kovalenko, J.W.F.~Valle and C.A.~Vaquera-Araujo,
  \emph{{Predictive Pati-Salam theory of fermion masses and mixing}},
  \href{https://doi.org/10.1007/JHEP07(2017)118}{\emph{JHEP} {\bfseries 07}
  (2017) 118} [\href{https://arxiv.org/abs/1705.06320}{{\ttfamily
  1705.06320}}].

\bibitem{deMedeirosVarzielas:2017sdv}
I.~de~Medeiros~Varzielas, G.G.~Ross and J.~Talbert, \emph{{A Unified Model of
  Quarks and Leptons with a Universal Texture Zero}},
  \href{https://doi.org/10.1007/JHEP03(2018)007}{\emph{JHEP} {\bfseries 03}
  (2018) 007} [\href{https://arxiv.org/abs/1710.01741}{{\ttfamily
  1710.01741}}].

\bibitem{Bernal:2017xat}
N.~Bernal, A.E.~C\'arcamo~Hern\'andez, I.~de~Medeiros~Varzielas and
  S.~Kovalenko, \emph{{Fermion masses and mixings and dark matter constraints
  in a model with radiative seesaw mechanism}},
  \href{https://doi.org/10.1007/JHEP05(2018)053}{\emph{JHEP} {\bfseries 05}
  (2018) 053} [\href{https://arxiv.org/abs/1712.02792}{{\ttfamily
  1712.02792}}].

\bibitem{deMedeirosVarzielas:2018vab}
I.~De~Medeiros~Varzielas, M.L.~L\'opez-Ib\'a\~nez, A.~Melis and O.~Vives,
  \emph{{Controlled flavor violation in the MSSM from a unified $\Delta(27)$
  flavor symmetry}}, \href{https://doi.org/10.1007/JHEP09(2018)047}{\emph{JHEP}
  {\bfseries 09} (2018) 047}
  [\href{https://arxiv.org/abs/1807.00860}{{\ttfamily 1807.00860}}].

\bibitem{CarcamoHernandez:2018djj}
A.E.~C\'arcamo~Hern\'andez, J.C.~G\'omez-Izquierdo, S.~Kovalenko and
  M.~Mondrag\'on, \emph{{$\Delta \left( 27\right)$ flavor singlet-triplet Higgs
  model for fermion masses and mixings}},
  \href{https://doi.org/10.1016/j.nuclphysb.2019.114688}{\emph{Nucl. Phys. B}
  {\bfseries 946} (2019) 114688}
  [\href{https://arxiv.org/abs/1810.01764}{{\ttfamily 1810.01764}}].

\bibitem{Bjorkeroth:2019csz}
F.~Bj\"orkeroth, I.~de~Medeiros~Varzielas, M.L.~L\'opez-Ib\'a\~nez, A.~Melis
  and O.~Vives, \emph{{Leptogenesis in $\Delta(27)$ with a Universal Texture
  Zero}}, \href{https://doi.org/10.1007/JHEP09(2019)050}{\emph{JHEP} {\bfseries
  09} (2019) 050} [\href{https://arxiv.org/abs/1904.10545}{{\ttfamily
  1904.10545}}].

\bibitem{Staub:2009bi}
F.~Staub, \emph{{From Superpotential to Model Files for FeynArts and
  CalcHep/CompHep}},
  \href{https://doi.org/10.1016/j.cpc.2010.01.011}{\emph{Comput. Phys. Commun.}
  {\bfseries 181} (2010) 1077}
  [\href{https://arxiv.org/abs/0909.2863}{{\ttfamily 0909.2863}}].

\bibitem{Staub:2010jh}
F.~Staub, \emph{{Automatic Calculation of supersymmetric Renormalization Group
  Equations and Self Energies}},
  \href{https://doi.org/10.1016/j.cpc.2010.11.030}{\emph{Comput. Phys. Commun.}
  {\bfseries 182} (2011) 808}
  [\href{https://arxiv.org/abs/1002.0840}{{\ttfamily 1002.0840}}].

\bibitem{Staub:2012pb}
F.~Staub, \emph{{SARAH 3.2: Dirac Gauginos, UFO output, and more}},
  \href{https://doi.org/10.1016/j.cpc.2013.02.019}{\emph{Comput. Phys. Commun.}
  {\bfseries 184} (2013) 1792}
  [\href{https://arxiv.org/abs/1207.0906}{{\ttfamily 1207.0906}}].

\bibitem{Staub:2013tta}
F.~Staub, \emph{{SARAH 4 : A tool for (not only SUSY) model builders}},
  \href{https://doi.org/10.1016/j.cpc.2014.02.018}{\emph{Comput. Phys. Commun.}
  {\bfseries 185} (2014) 1773}
  [\href{https://arxiv.org/abs/1309.7223}{{\ttfamily 1309.7223}}].

\bibitem{Varzielas:2016zjc}
I.~de~Medeiros~Varzielas, S.F.~King, C.~Luhn and T.~Neder, \emph{{CP-odd
  invariants for multi-Higgs models: applications with discrete symmetry}},
  \href{https://doi.org/10.1103/PhysRevD.94.056007}{\emph{Phys. Rev. D}
  {\bfseries 94} (2016) 056007}
  [\href{https://arxiv.org/abs/1603.06942}{{\ttfamily 1603.06942}}].

\bibitem{deMedeirosVarzielas:2017glw}
I.~de~Medeiros~Varzielas, S.F.~King, C.~Luhn and T.~Neder, \emph{{Minima of
  multi-Higgs potentials with triplets of $\Delta(3n^2)$ and $\Delta(6n^2)$}},
  \href{https://doi.org/10.1016/j.physletb.2017.11.005}{\emph{Phys. Lett. B}
  {\bfseries 775} (2017) 303}
  [\href{https://arxiv.org/abs/1704.06322}{{\ttfamily 1704.06322}}].

\bibitem{Ivanov:2014doa}
I.P.~Ivanov and C.C.~Nishi, \emph{{Symmetry breaking patterns in 3HDM}},
  \href{https://doi.org/10.1007/JHEP01(2015)021}{\emph{JHEP} {\bfseries 01}
  (2015) 021} [\href{https://arxiv.org/abs/1410.6139}{{\ttfamily 1410.6139}}].

\bibitem{Donoghue:1978cj}
J.F.~Donoghue and L.F.~Li, \emph{{Properties of Charged Higgs Bosons}},
  \href{https://doi.org/10.1103/PhysRevD.19.945}{\emph{Phys. Rev. D} {\bfseries
  19} (1979) 945}.

\bibitem{Georgi:1978ri}
H.~Georgi and D.V.~Nanopoulos, \emph{{Suppression of Flavor Changing Effects
  From Neutral Spinless Meson Exchange in Gauge Theories}},
  \href{https://doi.org/10.1016/0370-2693(79)90433-7}{\emph{Phys. Lett. B}
  {\bfseries 82} (1979) 95}.

\bibitem{Botella:2009pq}
F.J.~Botella, G.C.~Branco and M.N.~Rebelo, \emph{{Minimal Flavour Violation and
  Multi-Higgs Models}},
  \href{https://doi.org/10.1016/j.physletb.2010.03.014}{\emph{Phys. Lett. B}
  {\bfseries 687} (2010) 194}
  [\href{https://arxiv.org/abs/0911.1753}{{\ttfamily 0911.1753}}].

\bibitem{ParticleDataGroup:2020ssz}
{\scshape Particle Data Group} collaboration, \emph{{Review of Particle
  Physics}}, \href{https://doi.org/10.1093/ptep/ptaa104}{\emph{PTEP} {\bfseries
  2020} (2020) 083C01}.

\bibitem{Porod:2003um}
W.~Porod, \emph{{SPheno, a program for calculating supersymmetric spectra, SUSY
  particle decays and SUSY particle production at e+ e- colliders}},
  \href{https://doi.org/10.1016/S0010-4655(03)00222-4}{\emph{Comput. Phys.
  Commun.} {\bfseries 153} (2003) 275}
  [\href{https://arxiv.org/abs/hep-ph/0301101}{{\ttfamily hep-ph/0301101}}].

\bibitem{Porod:2011nf}
W.~Porod and F.~Staub, \emph{{SPheno 3.1: Extensions including flavour,
  CP-phases and models beyond the MSSM}},
  \href{https://doi.org/10.1016/j.cpc.2012.05.021}{\emph{Comput. Phys. Commun.}
  {\bfseries 183} (2012) 2458}
  [\href{https://arxiv.org/abs/1104.1573}{{\ttfamily 1104.1573}}].

\bibitem{Peskin:1990zt}
M.E.~Peskin and T.~Takeuchi, \emph{{A New constraint on a strongly interacting
  Higgs sector}}, \href{https://doi.org/10.1103/PhysRevLett.65.964}{\emph{Phys.
  Rev. Lett.} {\bfseries 65} (1990) 964}.

\bibitem{Marciano:1990dp}
W.J.~Marciano and J.L.~Rosner, \emph{{Atomic parity violation as a probe of new
  physics}}, \href{https://doi.org/10.1103/PhysRevLett.65.2963}{\emph{Phys.
  Rev. Lett.} {\bfseries 65} (1990) 2963}.

\bibitem{Kennedy:1990ib}
D.C.~Kennedy and P.~Langacker, \emph{{Precision electroweak experiments and
  heavy physics: A Global analysis}},
  \href{https://doi.org/10.1103/PhysRevLett.65.2967}{\emph{Phys. Rev. Lett.}
  {\bfseries 65} (1990) 2967}.

\bibitem{Goodsell:2018tti}
M.D.~Goodsell and F.~Staub, \emph{{Unitarity constraints on general scalar
  couplings with SARAH}},
  \href{https://doi.org/10.1140/epjc/s10052-018-6127-z}{\emph{Eur. Phys. J. C}
  {\bfseries 78} (2018) 649}
  [\href{https://arxiv.org/abs/1805.07306}{{\ttfamily 1805.07306}}].

\bibitem{Bechtle:2008jh}
P.~Bechtle, O.~Brein, S.~Heinemeyer, G.~Weiglein and K.E.~Williams,
  \emph{{HiggsBounds: Confronting Arbitrary Higgs Sectors with Exclusion Bounds
  from LEP and the Tevatron}},
  \href{https://doi.org/10.1016/j.cpc.2009.09.003}{\emph{Comput. Phys. Commun.}
  {\bfseries 181} (2010) 138}
  [\href{https://arxiv.org/abs/0811.4169}{{\ttfamily 0811.4169}}].

\bibitem{Bechtle:2013xfa}
P.~Bechtle, S.~Heinemeyer, O.~St\r{a}l, T.~Stefaniak and G.~Weiglein,
  \emph{{$HiggsSignals$: Confronting arbitrary Higgs sectors with measurements
  at the Tevatron and the LHC}},
  \href{https://doi.org/10.1140/epjc/s10052-013-2711-4}{\emph{Eur. Phys. J. C}
  {\bfseries 74} (2014) 2711}
  [\href{https://arxiv.org/abs/1305.1933}{{\ttfamily 1305.1933}}].

\bibitem{Haller:2018nnx}
J.~Haller, A.~Hoecker, R.~Kogler, K.~M\"onig, T.~Peiffer and J.~Stelzer,
  \emph{{Update of the global electroweak fit and constraints on
  two-Higgs-doublet models}},
  \href{https://doi.org/10.1140/epjc/s10052-018-6131-3}{\emph{Eur. Phys. J. C}
  {\bfseries 78} (2018) 675}
  [\href{https://arxiv.org/abs/1803.01853}{{\ttfamily 1803.01853}}].

\bibitem{ATLAS:2019old}
{\scshape ATLAS} collaboration, \emph{{Search for the Higgs boson decays $H \to
  ee$ and $H \to e\mu$ in $pp$ collisions at $\sqrt{s} = 13$ TeV with the ATLAS
  detector}}, \href{https://doi.org/10.1016/j.physletb.2019.135148}{\emph{Phys.
  Lett. B} {\bfseries 801} (2020) 135148}
  [\href{https://arxiv.org/abs/1909.10235}{{\ttfamily 1909.10235}}].

\bibitem{sarah:HB}
{Florian Staub, Mark Goodsell, Werner Porod}, \emph{SARAH Wiki}.
\newblock \url{https://gitlab.in2p3.fr/goodsell/sarah/-/wikis/HiggsBounds}.

\bibitem{ATLAS:2018gfm}
{\scshape ATLAS} collaboration, \emph{{Search for charged Higgs bosons decaying
  via $H^{\pm} \to \tau^{\pm}\nu_{\tau}$ in the $\tau$+jets and $\tau$+lepton
  final states with 36 fb$^{-1}$ of $pp$ collision data recorded at $\sqrt{s} =
  13$ TeV with the ATLAS experiment}},
  \href{https://doi.org/10.1007/JHEP09(2018)139}{\emph{JHEP} {\bfseries 09}
  (2018) 139} [\href{https://arxiv.org/abs/1807.07915}{{\ttfamily
  1807.07915}}].

\bibitem{Skands:2003cj}
P.Z.~Skands et~al., \emph{{SUSY Les Houches accord: Interfacing SUSY spectrum
  calculators, decay packages, and event generators}},
  \href{https://doi.org/10.1088/1126-6708/2004/07/036}{\emph{JHEP} {\bfseries
  07} (2004) 036} [\href{https://arxiv.org/abs/hep-ph/0311123}{{\ttfamily
  hep-ph/0311123}}].

\bibitem{Allanach:2008qq}
B.C.~Allanach et~al., \emph{{SUSY Les Houches Accord 2}},
  \href{https://doi.org/10.1016/j.cpc.2008.08.004}{\emph{Comput. Phys. Commun.}
  {\bfseries 180} (2009) 8} [\href{https://arxiv.org/abs/0801.0045}{{\ttfamily
  0801.0045}}].

\end{thebibliography}\endgroup
